\documentclass[aps,preprint,onecolumn,floatfix,superscriptaddress,nofootinbib]{revtex4-1}
\usepackage{booktabs}
\usepackage{dcolumn}

\AtBeginDocument{
\heavyrulewidth=.08em
\lightrulewidth=.05em
\cmidrulewidth=.03em
\belowrulesep=.65ex
\belowbottomsep=0pt
\aboverulesep=.4ex
\abovetopsep=0pt
\cmidrulesep=\doublerulesep
\cmidrulekern=.5em
\defaultaddspace=.5em
}

\newcolumntype{P}[1]{>{\centering\arraybackslash}p{#1}}

\newcommand{\lsim}{\mathrel{\mathop{\kern 0pt \rlap
  {\raise.2ex\hbox{$<$}}}
  \lower.9ex\hbox{\kern-.190em $\sim$}}}
\newcommand{\gsim}{\mathrel{\mathop{\kern 0pt \rlap
  {\raise.2ex\hbox{$>$}}}
  \lower.9ex\hbox{\kern-.190em $\sim$}}}

\newcommand{\gev}{\ensuremath{\,\mathrm{GeV}}}

\newcommand{\mchi}{\ensuremath{m_\chi}}
\newcommand{\sv}{\ensuremath{\langle\sigma v_{\rm rel}\rangle}}
\newcommand{\svnow}{\ensuremath{\langle\sigma v_{\rm rel}\rangle_0}}

\usepackage{amsmath}
\usepackage{graphicx,bm}
\usepackage[colorlinks,citecolor=blue]{hyperref}
\usepackage[caption=false]{subfig}
\usepackage{slashed}
\usepackage{setspace}
\usepackage[hang]{footmisc}
\usepackage{lipsum}

\begin{document}

\title{A convolutional-neural-network estimator of CMB constraints on dark matter energy injection}

\author{Wei-Chih Huang}
\email{huang@cp3.sdu.dk}
\affiliation{CP$^3$-Origins, University of Southern Denmark, Campusvej 55 5230 Odense M, Denmark}

\author{Jui-Lin Kuo}
\email{jui-lin.kuo@oeaw.ac.at}
\affiliation{Institute of High Energy Physics, Austrian Academy of Sciences, Nikolsdorfergasse 18, 1050 Vienna, Austria}

\author{Yue-Lin Sming Tsai}
\email{smingtsai@pmo.ac.cn}
\affiliation{Key Laboratory of Dark Matter and Space Astronomy, Purple Mountain Observatory, Chinese Academy of Sciences, Nanjing 210033, China}
\affiliation{Department of Physics, National Tsing Hua University,
Hsinchu 300, Taiwan}

\begin{abstract}

We show that the impact of energy injection by dark matter annihilation on the cosmic microwave background power spectra can be apprehended via a residual likelihood map.
By resorting to convolutional neural networks that can fully discover the underlying pattern of the map, we propose a novel way of constraining dark matter annihilation based on the Planck 2018 data.
We demonstrate that the trained neural network can efficiently predict the likelihood and accurately place bounds on the annihilation cross-section in a \textit{model-independent} fashion. The machinery will be made public in the near future.

\end{abstract}

\date{\today}    

\maketitle

\section{Introduction \label{sec:introduction}}

The nature of dark matter~(DM) remains one of the biggest unsolved enigmas in decades.
One simple, minimum assumption about DM is that DM couples to Standard Model~(SM) particles and its density follows
the thermal distribution.
In this scenario, thermal DM, when the temperature drops below the DM mass, the Boltzmann-suppressed DM-SM interactions can no longer keep up with the expansion of the universe and DM decouples from the thermal bath, aka freeze-out. 
To reproduce the correct density, the strength of the coupling should be similar to the weak interaction in the SM for the DM mass around the electroweak scale $\mathcal{O}(100\gev)$~\cite{Jungman:1995df,Bergstrom:2000pn,Bertone:2004pz}.
Such a coincidence is known as the ``WIMP miracle'' where WIMP standards for Weakly Interacting Massive Particle. 
To hunt for DM, various experiments have been developed, which can be categorized into direct detection, indirect detection, and searches at colliders.
Unfortunately, the long-awaited \textit{miracle} has not been observed so far,  
resulting in stringent limits, especially from direct detection, on the parameter space of the minimum DM theory.
To date, the DM mass for a successful freeze-out scenario is only allowed in some particular regions, 
such as a mass window of 
$\mathcal{O}({\rm MeV}\text{--}{\rm GeV})$~\cite{Pospelov:2007mp,DAgnolo:2015ujb,Matsumoto:2018acr}, 
mainly due to the kinetic threshold which leads to a sharp plummet in sensitivity toward the (sub-)GeV mass range in direct detection experiments considering nuclear recoil.
One can also avoid the stringent bounds by disconnecting DM freeze-out processes from those of detection, 
for example, with the help of the resonance enhancement~\cite{Matsumoto:2014rxa,Matsumoto:2016hbs,Athron:2018hpc,Bagnaschi:2019djj} 
or involving slightly heavier excited states into annihilation processes to increase the cross-section during the period of freeze-out, namely co-annihilation~\cite{Binetruy:1983jf,Griest:1990kh,Banerjee:2016hsk,Lu:2019lok}.

Besides the direct detection, the indirect searches for the consequent high-energy SM particles from DM annihilation are also of great interest.
In the early universe, these SM particles can interact and thermalize with the CMB photons, which leads to distortions on its black-body spectrum~\cite{Chluba:2011hw}.
Furthermore, in the post-recombination epoch, the injected energy can alter the evolution of the gas temperature and ionization fraction, in turn leaving imprints on the temperature and polarization power spectra of the CMB~\cite{Adams:1998nr,Chen:2003gz,Padmanabhan:2005es,Slatyer:2009yq}. 
It is worth mentioning that the CMB measurements are not subject to uncertainties on the local DM density and the kinetic threshold that brings down the sensitivity of direct detection.
As CMB experiments evolved and advanced significantly from the pre-WMAP epoch~\cite{Smoot:1992td,Netterfield:2001yq,Kovac:2002fg} to the WMAP~\cite{Bennett:2003ba} and the Planck satellite~\cite{Planck:2006aa}, we are now entering the precision era of CMB measurements. 
These data can provide independent, competitive bounds on the annihilation cross-section, especially on the (sub-)GeV mass range.
Note that other late-time indirect detection experiments 
such as Fermi-LAT~\cite{Ackermann:2015zua,Hoof:2018hyn,Oakes:2019ywx} 
and AMS-02~\cite{Boudaud:2016mos} also experience the loss of sensitivity~\cite{Roszkowski:2017nbc} 
toward the low-mass regime because of the instrumental threshold.   
All in all, the CMB provides a distinct, complementary probe into the nature of the thermal DM, especially important for low-mass DM and for models featuring $s$-wave annihilation that is not velocity suppressed.

The standard procedure to derive the CMB exclusion limit is first to translate the DM annihilation energy spectra 
into the evolution of the ionization fraction and gas temperature. Second, with the help of the Markov chain Monte Carlo~(MCMC) method, one maps out the likelihood distribution by including proper priors (if any) in the hyperspace consisting of both cosmological and nuisance parameters. Finally, the likelihood function can be projected onto the plane of the DM mass
and the cross-section by either marginalizing over~(Bayesian) or profiling out~(Likelihoodist) other parameters.
Then, the exclusion limit can be easily derived, \textit{e.g.}, by the $p$-value. 
The whole procedure, unfortunately, is very time-consuming and technically involved.
Besides, the task becomes unmanageable for scenarios where DM annihilates into various final states at the same
time. 
In this case, one has to generate the input annihilation spectrum for each set of the DM mass, cross-section and
branching ratios.

It has been pointed out in Refs.~\cite{Slatyer:2012yq,Madhavacheril:2013cna,Slatyer:2015kla,Slatyer:2015jla} 
that the DM-induced energy injection rate can be quantified by an \textit{effective parameter} 
$p_{\rm ann}=f (z)\sv/m_{\chi}$ where $f(z)$ is the energy deposition efficiency, $\sv$ is the thermal-averaged annihilation cross-section, and $m_{\chi}$ is the DM mass, respectively.
The dependency on the DM model is all encapsulated in $p_{\rm ann}$; as a result, $p_{\rm ann}$ can be directly constrained by the CMB measurements.
When the likelihood table as a function of $p_{\rm ann}$ is built via MCMC scans once and for all, 
one can simply derive constraints on $\sv$ for given $m_\chi$, if $f(z)$ is known; see Refs.~\cite{Slatyer:2015kla,Slatyer:2015jla} for derivation of $f(z)$.
This method is adopted by Planck collaboration~\cite{Ade:2015xua,Aghanim:2018eyx} where $f(z)$ is approximated as redshift-independent and only depends
on annihilation channels and the DM mass.
We note that Ref.~\cite{Slatyer:2016qyl} attempted to resolve the model-dependency of $f(z)$ by utilizing the principal component analysis to approximately parameterize the impact on the CMB power spectra with the DM-induced energy spectra. 
In this way, CMB constraints can be derived in a model-independent fashion\footnote{One still needs different treatments for DM decay, annihilation or other exotic processes such as a mixed channel considered in this work.}.

In this work, following the spirit of deriving CMB constraints in a \textit{model-independent} way discussed above, we use the perturbations on the evolution of the ionization fraction $dx_e^{\rm DM}/dz$ and gas temperature $dT_g^{\rm DM}/dz$, rather than $p_{\rm ann}$ or the energy spectra, to characterize the effects of DM-induced energy injection as they are more physically intuitive and general for any sort of energy injection.
In other words, we manage to build a likelihood map upon these quantities, 
with the aim to provide a \textit{single} tool of inferring the CMB constraints on generic models with
\textit{multiple} annihilation channels and \textit{arbitrary} branching ratios. 
That is, the tool can predict the value of likelihood regardless of details of DM models once the values of $dx_e^{\rm DM}/dz$ and $dT_g^{\rm DM}/dz$ are provided. 

Given the complexity of different final states and branching fractions, it is not practical to manually construct an enormous interpolation table to cope with a large parameter space.
We therefore resort to the power of Convolutional Neural Network~(CNN), to provide an very efficient way of placing constraints on DM models of \textit{multiple} annihilation channels with \textit{arbitrary} branching ratios.
As we have known, artificial intelligence~(AI) has been ubiquitous with numerous applications and has resulted in far-reaching influences in our daily lives.
Due to the availability of a tremendous amount of digital data, deep learning is a booming branch in AI, which has developed a large variety of neural networks and algorithms.
Out of many types of deep neural works, CNN is renowned for the capabilities of 
discovering patterns, shapes and correlations among input and output parameters~(especially powerful when the dimension of input parameters is large), and 
is employed here in light of the sequential nature of the input parameters, $dx_e^{\rm DM}/dz$ and 
$dT_g^{\rm DM}/dz$.
For a pedagogical introduction on CNN, see, \textit{e.g.}, Refs.~\cite{2015arXiv151108458O, 8308186}.
Furthermore, there have been quite a few applications of machine learning in cosmology and astronomy; see, \textit{e.g.}, Refs~\cite{Ball:2009wd,Ntampaka:2019udw, Carleo:2019ptp} for reviews. Deep neural networks have recently been utilized in the CMB physics such as
lensing reconstruction~\cite{Caldeira:2018ojb},
foreground modeling~\cite{Puglisi:2020deh, Farsian:2020adf, Petroff:2020fbf, Krachmalnicoff:2020rln} as well as data analysis~\cite{Krachmalnicoff2019, Sadr:2020rje}. 
A complete list of works that leverage machine learning to cosmology can be found
at the website\footnote{\url{https://github.com/georgestein/ml-in-cosmology}}.

Our workflow is sketched out as follows. First, we compute $dx_e^{\rm DM}/dz$ and $dT_g^{\rm DM}/dz$ following the methodology in Refs.~\cite{Kanzaki:2008qb,Kanzaki:2009hf,Kawasaki:2015peu}, and with those quantities we can attain the CMB angular power spectra via the Boltzmann solver \texttt{CLASS}~\cite{2011arXiv1104.2932L}.
Four typical annihilation channels are studied, $e^- e^+$, $b\bar{b}$, $W^- W^+$, $\mu^- \mu^+$, each with different combinations of
($m_\chi$, $\sv$).
Secondly, with the help of \texttt{MontePython}~\cite{Audren:2012wb,Brinckmann:2018cvx}, a parameter inference package for cosmology,  the power
spectra from \texttt{CLASS} are used to map out the likelihood in the hyperspace of cosmological
and nuisance parameters, given a set of the DM mass and cross-section.
We repeat this step for each set of ($m_\chi$, $\sv$).
Thirdly, for each ($m_\chi$, $\sv$) we profile out the cosmological and nuisance parameters by selecting the point with the maximum likelihood from the MCMC scans.
The corresponding likelihood and cosmological parameters are recorded.
Next, we train a CNN based on the input parameters, $dx_e^{\rm DM}/dz$ and $dT_g^{\rm DM}/dz$,
with the target variable\footnote{The target variable is the feature of a dataset on which the network make the prediction.} being the likelihood of the selected data points. 
Finally, we apply the trained network on unseen data generated from a new scenario with DM annihilating into all of the four channels mentioned above.

As we shall see below, the network yields consistent predictions for the new channel with those from MCMC scans. 
Similar performance can be expected even for completely new final states, as long as their energy spectra are not contrasting with those seen by the network. 
In addition, the model is very scalable to different
scenarios; for instance, by adding data points from decaying DM models to the training dataset,
the network can handle both annihilating and decaying cases without extra tweaks -- more training data of different types, better performance and broader applicability.
Furthermore, one can train the network to predict
not only the likelihood but also the corresponding best-fit values of the cosmological parameters by
simply including these parameters as the target variables.

The rest of the paper is organized as follows.
In Sec.~\ref{sec:CMB}, we begin by summarizing the calculation of the ionization fraction and gas temperature evolution in the presence of DM energy injection. Then, we demonstrate via a residual likelihood map that the DM-induced changes on the ionization fraction and gas temperature can closely characterize the effects on the CMB.
We furthermore explain how the CMB power spectra are computed and how MCMC scans over the parameter space are performed with the existing tools. 
In Sec.~\ref{sec:ML}, we discuss the network structure and data training. 
In Sec.~\ref{sec:result}, the good performance of the network in terms of the accuracy of
predictions and the CMB exclusion limits is displayed, and the decent performance happens to unseen data as well.
Finally, we conclude in Sec.~\ref{sec:conclusion}.


\section{CMB constraints on energy injection from DM}
\label{sec:CMB}

DM signals have not been detected by the measurement of the CMB anisotropy thus far, which constrains the energy injection sourced through DM to be small, compared to that of the standard cosmology and statistical uncertainties.
In this section, we start with elaborating on how we calculate the effect of DM energy deposition on the ionization fraction and gas temperature evolution.
Then we show the statistical importance of DM contributions at different redshifts can be captured by a residual likelihood map in a model-independent way.
Finally, we detail the conventional global fitting approach that utilizes the MCMC method to scan over a hyperparameter space of the cosmological parameters given DM model parameters to obtain robust CMB constraints on DM annihilation.

\subsection{DM contribution to ionization and heating}
\label{sec:stdFormula}

\begin{figure*}[t]
  \centering
\includegraphics[width=0.49\textwidth]{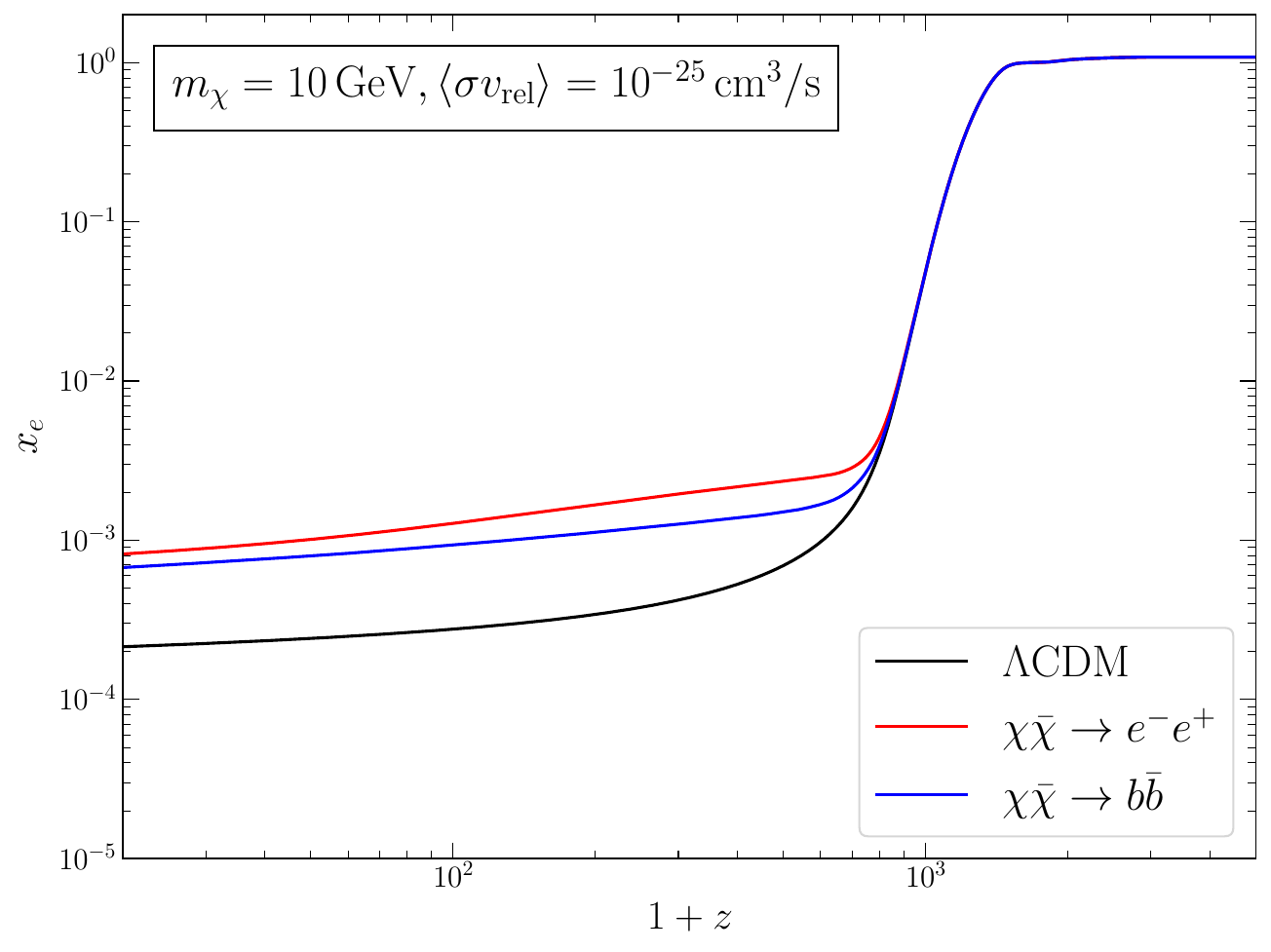}
\includegraphics[width=0.49\textwidth]{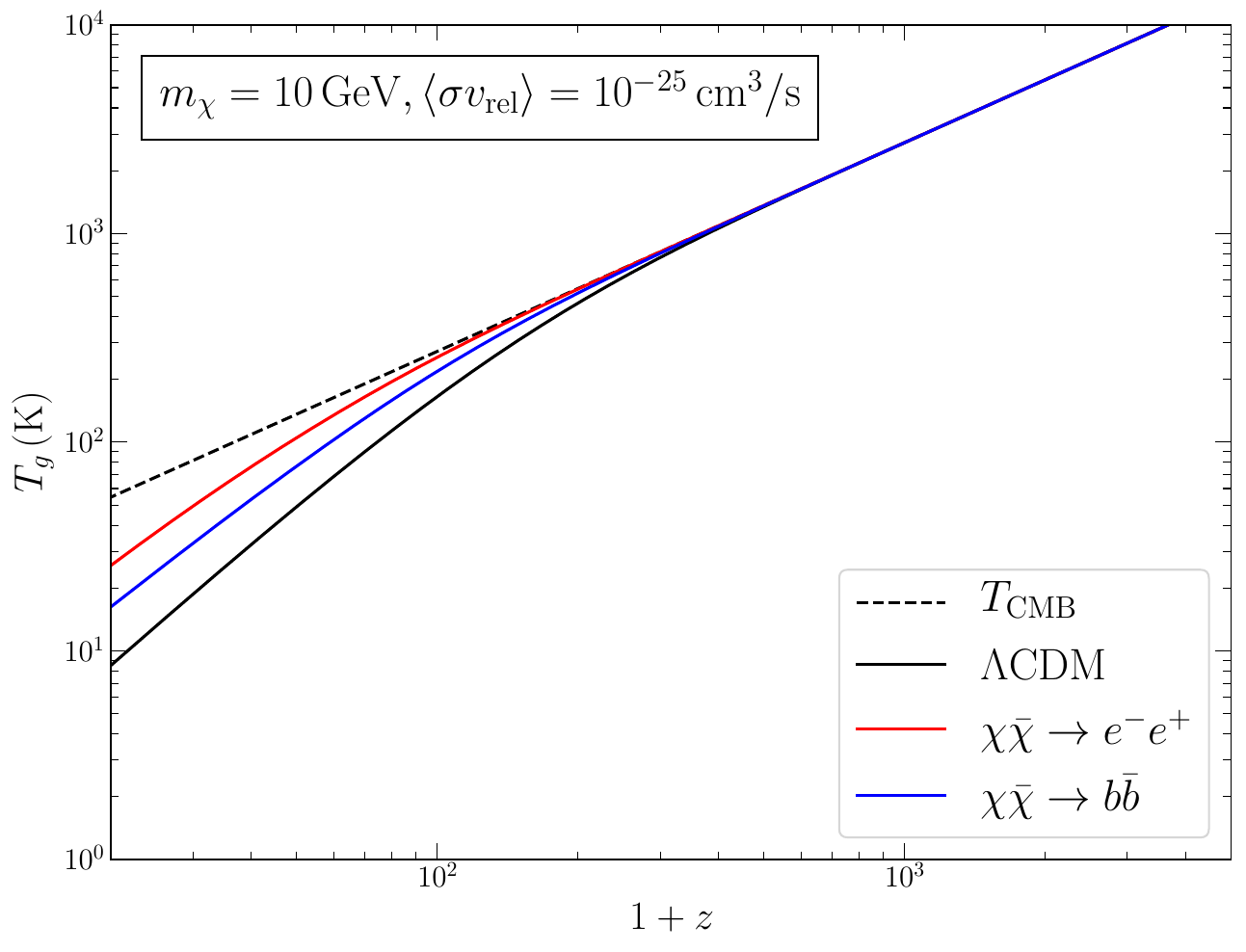}
\caption{Exemplary evolutions of ionization fraction (left panel) and gas temperature (right panel) for both $\Lambda$CDM and cases with the energy injection from DM annihilation. 
The benchmark values of DM parameters are $m_\chi = 10\,{\rm GeV}$ and $\sv = 10^{-25}\,{\rm cm}^3/{\rm s}$.
The $\Lambda$CDM scenario, the $e^- e^+$ and $b\bar{b}$ channels are shown by black, red and blue curves, respectively. 
We can clearly see the additional ionization and heating caused by DM annihilation.
}
  \label{Fig:thermalhistory}
\end{figure*}

For calculation of DM-induced effects on the evolution of the ionization fraction $x_e$ and the gas temperature $T_g$, we solely consider $s$-wave annihilation as the cross-section is not suppressed by the DM relative velocity $v_{\rm rel}$.  
The DM contribution to ionization and heating can be written as~\cite{Kanzaki:2008qb,Kanzaki:2009hf,Kawasaki:2015peu,Cheung:2018vww}
\begin{align}
\label{Eq:energy_injection_s-wave}
-\left[{\frac{dx_e^{\rm DM}}{dz}}\right]_{s\text{-}{\rm wave}} &= \displaystyle\sum_{F} {\rm Br}_F
        \int_z \dfrac{dz'}{H(z')(1+z')} \dfrac{n^2_\chi(z')\sv}{2n_{\rm H}(z')} 
\dfrac{m_\chi}{E_{\rm RY}} \dfrac{d\chi^F_i(m_\chi,z,z')}{dz}\,, \nonumber \\
-\left[ \dfrac{dT_g^{\rm DM}}{dz}\right]_{s\text{-}{\rm wave}} &= \displaystyle\sum_{F} {\rm Br}_F
        \int_z \dfrac{dz'}{H(z')(1+z')}\dfrac{n^2_\chi(z'){\sv}}{3n_{\rm H}(z')} 
        m_\chi \dfrac{d\chi^F_h(m_\chi,z,z')}{dz}\,,
\end{align}
where $H(z)$, $n_\chi(z)$ and $n_{\rm H} (z)$ are the Hubble parameter, DM number density, 
and hydrogen atom number density, respectively, as functions of redshift $z$.
Notice that $z'$ stands for the redshift at which the energy is injected, while $z$ is the redshift when the injected energy is absorbed. In this way, the differential ionization fraction and gas temperature
at redshift $z$ account for accumulative effects from all energy deposition that happened at earlier redshifts, $z' > z$.
The Rydberg energy $E_{\rm RY} \equiv 13.6\,{\rm eV}$ is the threshold energy for ionizing a hydrogen atom.
The notation $F$ denotes different final states with branching ratio ${\rm Br}_F$. 
The terms $d\chi_{i,h}^F (m_\chi,z,z')/dz$ in Eq.~\eqref{Eq:energy_injection_s-wave}, 
which represent the fraction of injected energy going into ionization ($i$) and heating ($h$), are given by
\begin{equation}
\label{Eq:fraction_inj_E}
    \dfrac{d\chi_{i,h}^F (m_\chi,z,z')}{dz} = \int dE \, \dfrac{E}{m_\chi} \left[2 \dfrac{dN_e^F}{dE} \dfrac{d\chi_{i,h}^e (E,z,z')}{dz}+\dfrac{dN_\gamma^F}{dE} \dfrac{d\chi_{i,h}^\gamma (E,z,z')}{dz}\right]\,.
\end{equation}
%
Calculation of the fractions  $d\chi_{i,h}^{e,\gamma} (E,z,z')/dz$ is explained in Refs.~\cite{Kanzaki:2008qb,Kanzaki:2009hf,Kawasaki:2015peu} 
and the final energy spectra of electrons and photons per DM annihilation $dN_{e,\gamma}^F /dE$ are obtained from
the package \texttt{LikeDM}~\cite{Huang:2016pxg}. It interpolates 
three dimensional \texttt{PPPC4}~\cite{Cirelli:2010xx} tables\footnote{There exist alternative tables in Ref.~\cite{Amoroso:2018qga} which employed a newer version \texttt{Pythia} 8.2~\cite{Sjostrand:2014zea}.} which take into account electroweak corrections computed in Ref.~\cite{Ciafaloni:2010ti}.

We should point out that  
the existing CMB constraints~\cite{Kawasaki:2015peu} derived based on Eq.~\eqref{Eq:energy_injection_s-wave} are in accord with results of other studies~\cite{Slatyer:2009yq,Slatyer:2012yq,Slatyer:2015jla,Slatyer:2015kla} that employ a different method of computing the DM energy injection.
%
Moreover, the cosmological boost factor discussed in Ref.~\cite{Cheung:2018vww} is not considered here as the epoch of interest is still in the linear regime.
In Fig.~\ref{Fig:thermalhistory}, we demonstrate the evolution of $x_e$ and $T_g$ for $s$-wave annihilation into $e^- e^+$ and $b\bar{b}$ with $m_\chi = 10\,{\rm GeV}$ and $\sv = 10^{-25} \,{\rm cm^3}/{\rm s}$.

In general, the energy injection rate of DM  annihilation 
is proportional to the velocity-averaged cross-section $\sv$ multiplied by the DM number density squared $n_\chi^2$. 
One can simply Taylor expand $\sv$ in powers of  $v_{\rm rel}$. Only the velocity-independent component ($s$-wave) is explored here, but there exist 
$v_{\rm rel}^2$-dependent component ($p$-wave)~\cite{Diamanti:2013bia,Liu:2016cnk,An:2016kie} and annihilation through resonance (Breit-Wigner enhancement)~\cite{Ibe:2008ye,Guo:2009aj,Bi:2011qm}, which are also phenomenologically interesting.
However, we do not explore these scenarios as they are either subdominant due to velocity suppression or more complicated in light of additional model parameters such as the mass and width of the intermediate particle.

Apart from annihilation, DM can inject energy into the thermal plasma by decaying into SM particles.
In this case, the energy injection rate of DM decay is proportional to $n_\chi$~(scales as $(1+z)^{3}$), which results in $dx_e^{\rm DM}/dz \propto (1+z)^{-5/2}$
as opposed to $dx_e^{\rm DM}/dz \propto (1+z)^{1/2}$ for $s$-wave annihilating DM.
Therefore, for decaying DM, energy deposition  at low redshifts becomes more important than that at high redshifts. 
For the mass region of GeV--TeV considered in this study, constraints on decaying DM from indirect search~\cite{Yuksel:2007dr,PalomaresRuiz:2007ry,Zhang:2009ut,Cirelli:2009dv,Bell:2010fk,Dugger:2010ys,Cirelli:2012ut,Murase:2012xs,Essig:2013goa,Mambrini:2015sia} are more stringent than the CMB bound~\cite{Slatyer:2016qyl}.
Nevertheless, the constraints from CMB and reionization~\cite{Liu:2016cnk} are still crucial for sub-GeV DM decay,
which will be pursued in the future.

As a final remark of this section, although we only consider $s$-wave annihilation, we note that the methodology of computing the differentials of $x_e$ and $T_g$ described in Refs.~\cite{Kanzaki:2008qb,Kanzaki:2009hf,Kawasaki:2015peu} can be applied to all the different scenarios mentioned above.
For completeness, the derivations of formulas similar to Eq.~\eqref{Eq:energy_injection_s-wave} are given in App.~\ref{Sec:decay} for DM decay, in App.~\ref{Sec:pwave} for $p$-wave annihilation, and in App.~\ref{Sec:resonance} for annihilation via resonance.

\subsection{CMB residual likelihood map}
\label{sec:resid}

From a residual likelihood map\footnote{The term ``residual'' refers to the deviation from the background value.}, one can easily see at which redshift intervals the changes in the evolution of $x_e$ and $T_g$ lead to significant effects on the CMB power spectra. 
To some extent, the map represents the underlying pattern the neural network will discover during the training process.
In terms of model-independence, we find that
the quantities $dx_e^{\rm DM}/dz$ and $dT_g^{\rm DM}/dz$ are more suited to represent the DM impact
than the parameters of DM models like the mass and cross-section.
We assume the DM contribution is only perturbation to the standard cosmology, which is well-justified for the parameter space close to the exclusion limit, where the effect on the CMB power spectra is linearly proportional to $dx^{\rm DM}_e/dz$ and $dT^{\rm DM}_g/dz$. 
To demonstrate the dependency, we further ignore $dT^{\rm DM}_g/dz$ in constructing the residual likelihood map because its effect on the CMB power spectra is rather minor compared to that of $dx_e^{\rm DM}/dz$, simplifying the map from three dimensions ($z$, $dx^{\rm DM}_e/dz$, $dT^{\rm DM}_g/dz$) to two dimensions ($z$, $dx^{\rm DM}_e/dz$).

In general, the statistics strength $\chi^2_{\rm test}(m_\chi, \sv) = -2\ln\mathcal{L}(m_\chi, \sv)$ is computed by contrasting the expected CMB angular power spectra given a model
to the observed one.
For a robust derivation of $\chi^2_{\rm test}$, we need to perform the full numerical calculation of CMB power spectra to obtain the likelihood for each set of the DM parameters, \textit{e.g.}, through a MCMC scan.
The results from the full numerical calculation are presumably more accurate but very time-consuming. 
For demonstration, we simply reconstruct the likelihood in terms of $dx^{\rm DM}_e/dz$ to generate a residual likelihood map based on $dx^{\rm DM}_e/dz$.
First, we discretize a given $dx^{\rm DM}_e/dz$ curve into $n$ log-spaced bins in redshifts.
The assumption of small DM contributions allows us to decompose the summed statistical strength by
\begin{equation}
\delta \chi^2 \approx \sum_{i,j} C_{ij} \sqrt{\delta\chi_i^2 \delta\chi_j^2}, 
\label{eq:final_chi2}
\end{equation}
where $C_{ij}$ is the element of the covariance matrix $C$ with $i$ and $j$ representing the $i$-th and
$j$-th bin. 
Here, we define 
\begin{equation}
\delta\chi^2 \equiv \chi_{\rm test}^2 -\chi_{\Lambda{\rm CDM}}^2 \,,
\label{eq:chisq}
\end{equation}
where $\chi^2_{\Lambda{\rm CDM}}$ is the statistics strength for the standard $\Lambda$CDM cosmology, and $\delta \chi_i^2 = \chi_i^2 - \chi^2_{\Lambda{\rm CDM}}$.
The individual $\chi_{i}^2 = -2 \ln \mathcal{L}_i$ can be obtained by performing a full numerical calculation of CMB power spectra, with DM contribution being a kernel function 
\begin{equation}
\left[\frac{dx_e^{\rm DM}}{dz}\right]_i =  
 \left\{
        \begin{array}{ll}
            \mathcal{N}_{\rm DM}, & \quad  {\rm if} ~~ z_i\leq z \leq z_{i+1}, \\
            0, & \quad {\rm else.}
        \end{array}
    \right.
      \label{eq:kernel}
\end{equation}
Here $\mathcal{N}_{\rm DM}$ is treated as a model-independent parameter. 
For any given DM annihilation channel, the values of $\mathcal{N}_{\rm DM}(z)$ can be computed using Eq.~\eqref{Eq:energy_injection_s-wave} for each redshift bin.
We tabulate $\mathcal{N}_{\rm DM}$ and their resulting $\chi_{i}^2$ for later usage.  
In case of no correlation among the bins, the value of $C_{ij}$ is just a normalization factor $1/n$.
By contrast, the components can be negative for $i\neq j$ in the presence of correlation.
For simplicity, here we assume $C_{ij} = 0$ for $i\neq j$,
which is adequate as long as the DM contribution is small.
The correlation will automatically be taken into account in the MCMC scans.

\begin{figure*}[t]
  \centering
\includegraphics[width=0.75\textwidth]{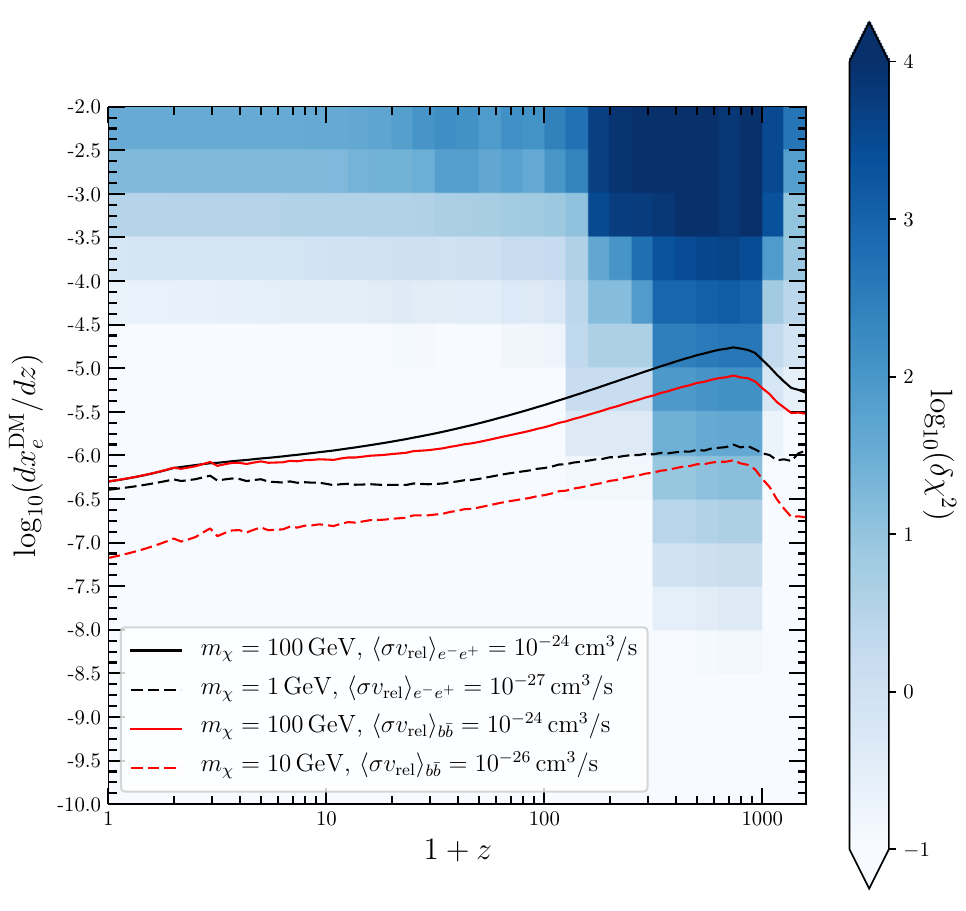}
\caption{The residual CMB likelihood map $\delta \chi^2 (z, dx_e^{\rm DM}/dz)$. We also plot the $dx_e^{\rm DM}/dz$ curves for $e^- e^+$ and $b\bar{b}$ channels with various choices of $(m_\chi, \sv)$ for comparison.
The energy injection from DM annihilation can significantly affect the CMB power spectra in $z = [600, 1000]$. 
}
  \label{like_map}
\end{figure*}
The residual likelihood map for the individual redshift bins is presented in Fig.~\ref{like_map} together with exemplary $dx_e^{\rm DM}/dz$ curves from different DM annihilation channels.
As indicated by regions with a darker color in Fig.~\ref{like_map}, the CMB power spectra are more sensitive to the energy injection in the redshift range $[600, 1000]$. Our findings are in agreement with the observations in Refs.~\cite{Slatyer:2009yq,Slatyer:2012yq,Slatyer:2015jla,Slatyer:2015kla}.
Besides, this map demonstrates that the derived quantity $dx_e^{\rm DM}/dz$ is indeed a suitable quantity for delineating, in a model-independent fashion, the effect of an arbitrary DM model on the CMB.
In summary, we would like to emphasize that the residual likelihood map 
is only used to demonstrate: 
i) the important redshift-interval for CMB constraints on DM energy injection is $z=[600,1000]$ and 
ii) $dx_e^{\rm DM}/dz$ and $dT_g^{\rm DM}/dz$ prove capable of characterizing the DM impacts and
hence are proper input parameters for neural networks.

\subsection{Cosmological scans for DM annihilation}
\label{sec:global_fitting}

Although the simplified CMB residual likelihood map can depict the DM impacts, it is still crucial to consider the  correlation among energy injections from different redshifts to infer a robust, precise CMB constraint.
The standard procedure of achieving that is to perform a global fitting of DM models as follows.  
First, we calculate $dx_e^{\rm DM} /dz$ and  $dT^{\rm DM}_g /dz$ according to the formulas in Sec.~\ref{sec:stdFormula}. 
The tabulated $dx^{\rm DM}_e /dz$ and $dT^{\rm DM}_g /dz$ 
are then inserted to a Boltzmann equation solver, \textit{e.g.}, \texttt{CAMB}~\cite{Lewis:1999bs,Howlett:2012mh} or \texttt{CLASS}~\cite{2011arXiv1104.2932L}, to obtain the CMB angular power spectra. 
We calculate the likelihood based on the resulting CMB angular power spectra and the Planck 2018 data~\cite{Aghanim:2019ame}, out of which we \textit{exclusively} involve the baseline high-$\ell$ power spectra (\texttt{TT}, \texttt{TE}, and \texttt{TE}), low-$\ell$ power spectrum \texttt{TT}, low-$\ell$ HFI polarization power spectrum \texttt{EE}, and lensing power spectrum \texttt{lensing}.

In practice, to determine the proper statistical strength with the systematic uncertainties of $\Lambda$CDM 
for each of ($m_\chi$, $\sv$) sets, one has to simultaneously scan over all cosmological parameters
$\mathcal{C} = (\Omega_b h^2,\, \Omega_\chi h^2,\, 100 \, \theta_s,\, h,\, \ln 10^{10}A_s,\, n_s,\, \tau)$ as well as any additional nuisance parameters $\mathcal{N}$ associated with the data sets involved.
The cosmological parameters are the reduced Hubble constant $h$, baryon density parameter $\Omega_b$, DM density parameter $\Omega_\chi$, the angle subtended by the sound horizon $\theta_s$, the primordial curvature power spectrum $A_s$ at $k_0 = 0.05\,{\rm Mpc}^{-1}$, the scalar spectrum power-law index $n_s$, and the Thomson scattering optical depth due to reionization $\tau$, respectively.
In this work, we utilize \texttt{MontePython}~\cite{Audren:2012wb,Brinckmann:2018cvx} to compute 
the likelihood probability density $\mathcal{L}_i(m_\chi, \sv, \mathcal{C}, \mathcal{N})$ 
with the CMB power spectra computed by \texttt{CLASS}, and scan the space of cosmological and nuisance parameters via a MCMC method, Metropolis-Hastings. Note that scans over the nuisance parameters are handled differently from the cosmological ones according to the method of \textit{fast sampling}~\cite{Brinckmann:2018cvx} as jumps along the former need not involve the time-consuming Boltzmann code.

For each scan of ($m_\chi$, $\sv$), we profile out the cosmological and nuisance parameters by choosing a single point with the maximum likelihood, whose value is recorded together with the corresponding cosmological parameters, namely
\begin{equation}
\mathcal{L}(m_\chi, \sv) = \max_{\mathcal{C}, \, \mathcal{N}} \left[\mathcal{L}_i(m_\chi, \sv, \mathcal{C}, \mathcal{N})\right].
\label{eq:max_like}
\end{equation}
In this work, $\chi^2_{\rm test}(m_\chi, \sv)$ in Eq.~\eqref{eq:chisq} is defined as 
$-2\ln\mathcal{L}(m_\chi, \sv)$, which is legitimate as $\mathcal{L}_i(m_\chi, \sv, \mathcal{C},\mathcal{N})$ 
on the right-hand side
of Eq.~\eqref{eq:max_like} are well-approximated by the Gaussian distribution.
Moreover, our statistic strength $\delta\chi$ in Eq.~\eqref{eq:chisq} 
can be understood as the null-signal approach. 
Therefore, the $95\%$ confidence level for $\sv$ given $m_\chi$ corresponds to 
$\delta\chi^2=2.71$, 
assuming a one-side $\chi^2$-distribution.
The derived constraints from the MCMC scans will be discussed later in Sec.~\ref{sec:result} together with the network performance.
%

\section{Network architecture and data training}\label{sec:ML}

In this section, we detail how we choose the network structure as well as the training procedures.
We utilize \texttt{TensorFlow}~\cite{TensorFlow} which is an end-to-end, open-source machine learning platform that
has built-in \texttt{Keras}~\cite{Keras}, a deep learning application programming interface written in \texttt{Python}.
As demonstrated in Sec.~\ref{sec:resid}, the evolution of ionization fraction $x_e$ and gas temperature $T_g$ are capable of characterizing, in a model-independent way, impacts of the DM-induced energy injection into the thermal plasma
-- larger deviation from the predictions of standard cosmology, larger $\chi^2_{\rm test}$ from CMB angular power spectra and hence more likely being excluded.
Consequently, $dx_e^{\rm DM}/dz$ and $dT_g^{\rm DM}/dz$
are chosen as the input parameters instead of the DM mass, cross-section, and annihilation channel.
In this way, we \textit{assist networks with the knowledge of physics} by choosing proper input parameters. 

For the format of input parameters, each curve of $\log_{10}(dx_e^{\rm DM}/dz)$ and $\log_{10}(dT_g^{\rm DM}/dz)$ are discretized into 30 equally-spaced points in the log-scale of $(1+z)$. The reason why
we use the logarithm is that values of the differentials vary over many orders of magnitude. Taking their exponents as the input parameters significantly reduces the parameter range, which dramatically expedites 
the network's optimization algorithm. 
In this approximation, the continuous values of the differentials are represented by the values of individual points in these 30 bins.
We then preprocess the data by calculating 
the standard score~($z$-score) for the whole data set in each of the 30 bins. Namely, $x^j_i \to (x^j_i - \langle x^j \rangle)/\sigma_{x^j}$, where $x^j$ denotes all data points located in the $j$-th bin while $\langle x^j \rangle$ and $\sigma_{x^j}$ correspond to the mean and the standard deviation of points ${x^j_i}$, respectively. All in all,
for each data point, the input parameter is a two-dimensional array with a shape of $(30,2)$.

The output parameter~(aka the target variable in machine learning) for the network 
is simply the minimum value of $\chi^2_{\rm test}$ obtained by 
full MCMC scans from \texttt{MontePython}, given the DM mass and annihilation cross-section. 
Note that we do not use the simplified residual likelihood map discussed in the previous section at all.
In practice, $ \delta \chi^2 = \chi^2_{\rm test} - \chi^2_{\Lambda\rm{CDM}}$~($\ll \chi^2_{\rm test}$) is used instead, to facilitate network optimization such that layers with small values of weights
suffice for accurate predictions. Otherwise, the optimization algorithm has to spend more time searching in a larger parameter space.
Note that \texttt{MontePython} also yields best-fit values for the cosmological parameters $\mathcal{C}$. In principle, one can train networks to make predictions on all the best-fit values provided by \texttt{MontePython}\footnote{One can simply construct an individual network for each of the cosmological parameters with the same procedure described here. Alternatively, one can contrive a more complicated network architecture for simultaneous predictions on all the parameters.}.
The best-fit values of the cosmological parameters across the dataset, however, are quite centralized with tiny variation; thus, we concentrate on the prediction of $\delta \chi^2$.

\begin{figure}
\begin{center}
\includegraphics[width=0.9\textwidth]{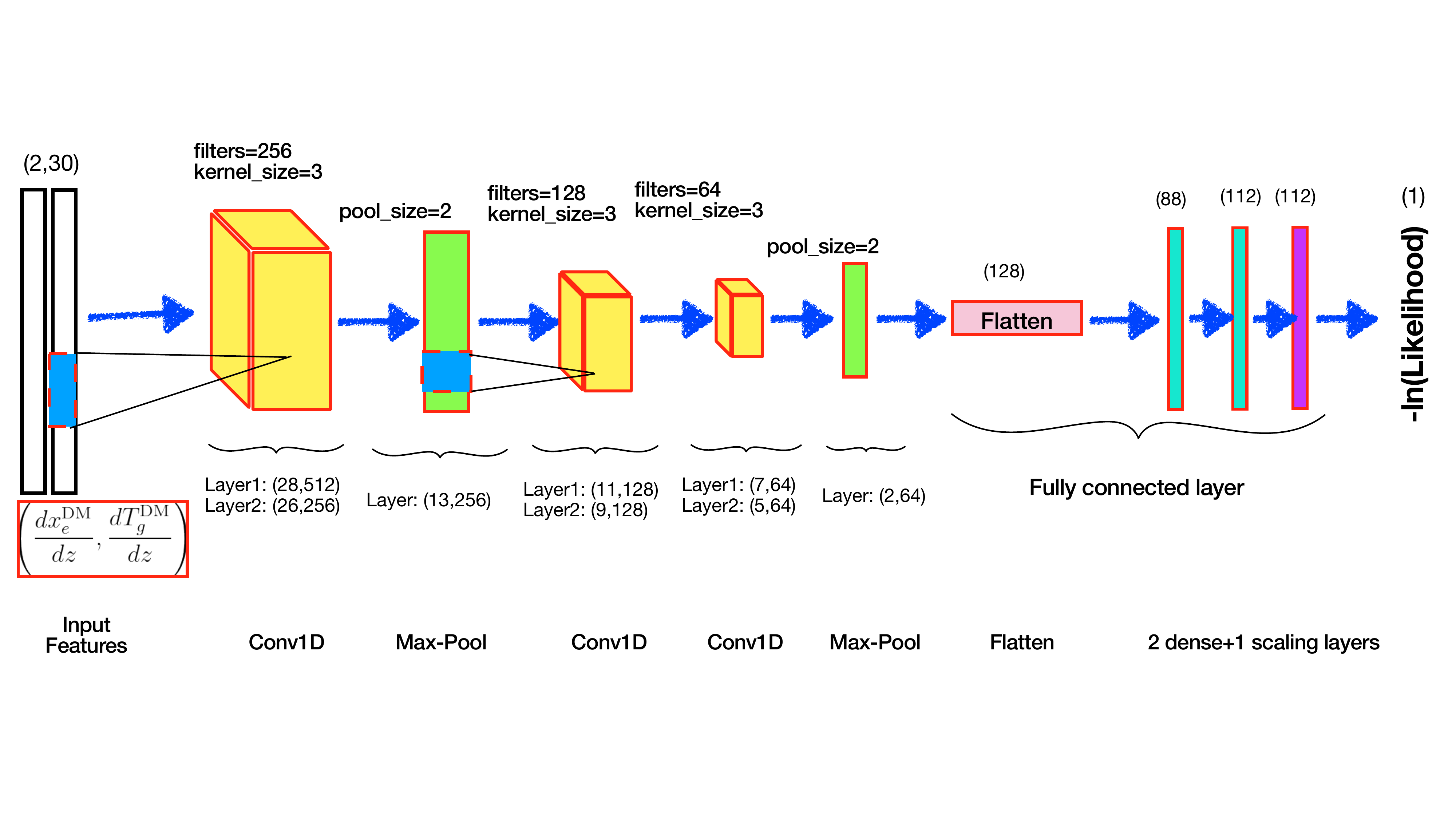}
\caption{The architecture of our machine. 
See main text for more details.
The output shape of the layer is specified by the values in the parentheses
which is also the input shape of the next layer.
The structure of networks is analogous to a previous work~\cite{Tsai:2020vcx} 
which deals with cosmic ray spectra in a similar manner.
\label{Fig:Architecture}}
\end{center}
\end{figure}

We are now in a position to discuss the network structure.
In light of the redshift dependence of the $dx_e^{\rm DM}/dz$ and $dT_g^{\rm DM}/dz$, the sequence and correlation of the input parameters and matters. That is, swamping values of different redshifts represent different physical scenarios.
As generic deep neural networks will not be able to properly capture the nature of the sequence, one-dimensional CNNs, \texttt{Conv1D} layers, are employed to extract correlation and features of the input parameters as shown in Fig.~\ref{Fig:Architecture} that pictorially displays the network architecture and exhibits the evolution of the input array's shape through different layers.
Applying a filter of size 3~(with 2 channels corresponding to the depth of $2$ in the input array for $dx_e^{\rm DM}/dz$ and $dT_g^{\rm DM}/dz$ respectively) converts an input array of dimension $(30, 2)$ into an array of $(28, 1)$.
An input array is thus transferred into an array of $(28, 512)$ with the first \texttt{Conv1D} layer of 512 filters.
The \textit{pooling layers}, \texttt{MaxPooling1D} are deployed among \texttt{Conv1D} layers to enhance extracted features by choosing the maximum values within the window of pooling kernels, at the same time reducing the size of the outputs from \texttt{Conv1D} layers. We refer readers to websites of \texttt{Keras} or \texttt{TensorFlow} for technical details of CNNs and other types of layers used in this work.

Finally, the convoluted 2-D array is flattened into a 1-D array and fed into two \textit{fully-connected layers},
\texttt{Dense}, with 88 and 112 neurons, respectively.
The output layer consists of a single neuron for predicting the value of $\delta \chi^2$. 
The second to last layer is the \texttt{Lambda} layer, which simply rescales the output of the previous layer by a factor of the maximum value of $\delta \chi^2$ of the entire data set. The insertion of the \texttt{Lambda} layer follows the same logic of using
$\delta \chi^2$ as the target variable -- confining searches of minimization to a small parameter space accelerates the optimization process.
Furthermore, the number of neurons for the two fully-connected layers are determined by \texttt{Keras Tuner}~\cite{kt_tuner} that enables search for the best set of hyperparameters\footnote{Hyperparameters, different from parameters of a network, \textit{e.g.}, weights of hidden layers, are those related to network architecture, such as the number and width of hidden layers, and to the learning algorithm, such as the learning rate for gradient descent.} of networks automatically and efficiently.
There are few available algorithms of searching for the best hyperparameters in \texttt{Keras}.
We choose the \texttt{Hyperband} tuning algorithm~\cite{Hyperband}; see, \textit{e.g.}, Ref.~\cite{kt_tuner_tf} for more details and examples.
The algorithm uses adaptive resource allocation and early-stopping to quickly pin down a model of high-performing. This is carried out via a sports championship style bracket. It trains a large number of models for a few epochs and moves forward only the top-performing half of models to the next round
until the specified \texttt{max\_epochs} is reached.
The activation function of \texttt{ReLU}, \textit{i.e.}, $x \to x~(x \to 0)$ for $x>0~(x\leq 0)$, is used for the \texttt{Conv1D} and \texttt{Dense} layers while the output layer assumes the linear activation function~($x \to x$).

For the training procedure, the total 39421 points from four independent channels  $\chi\bar{\chi} \to e^-e^+$, $b\bar{b}$,
 $W^-W^+$ and $\mu^-\mu^+$ -- each data point corresponds to a set of $(m_\chi, \sv)$ and the corresponding $\delta \chi^2$ -- are randomly shuffled and split into a training set~($80\%$ of 39421 points) and a validation set~($20\%$).
 The 39421 points come from grid scans on the plane of $(m_\chi, \sv)$ in the four channels. Since the focus of this work is for the neural network to quickly predict whether or not the induced energy injection from a given DM model is excluded by the CMB constraints, the regions of interest for the grid scans are confined to be $ 10^{-3}~\sv_{95} \leq \sv \leq 20~\sv_{95}$,
 given a DM mass and annihilation channel. The symbol $\sv_{95}$ represents the existing bound at $95\%$ confidence level on the annihilation cross-section. In this way, the network is able to reproduce
 the bound $\sv_{95}$ efficiently and precisely.
As we shall see below, the evaluation of the trained network performance is carried out with unseen data from a mixed
channel where DM particles annihilate into all of the four final states.
We choose the mean squared error for the loss function, which quantifies the difference between the network prediction
and value of the target variable.
In addition, the root mean squared error as the metrics is used to monitor the network performance during the training.  The Adam algorithm\footnote{Adam optimization~\cite{2014arXiv1412.6980K} is a method of stochastic gradient descent according to the adaptive estimation of first-order and second-order moments.} is adopted to minimize the loss function with a learning rate of $10^{-4}$.

\begin{figure}
\begin{center}
\includegraphics[width=0.6\textwidth]{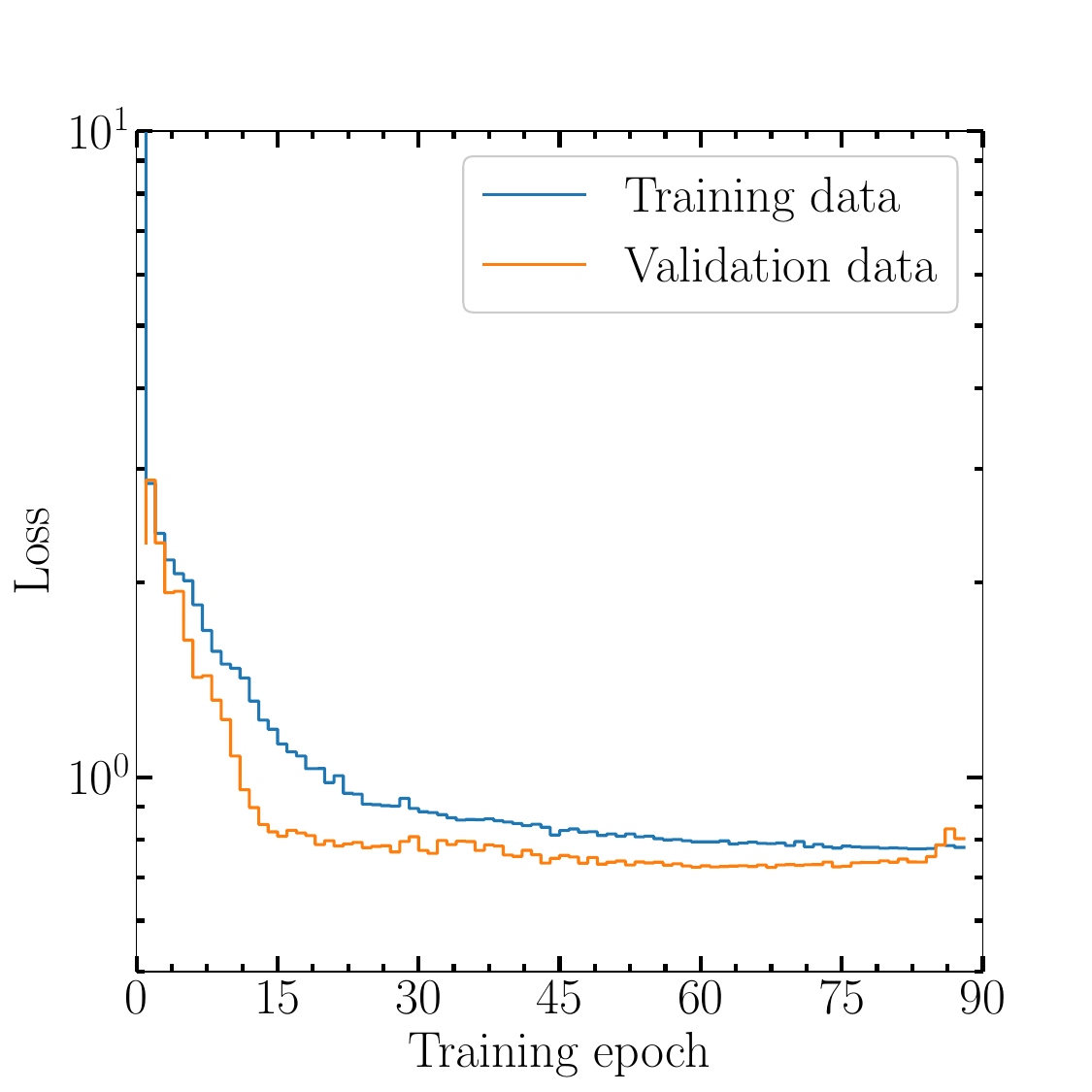}
\caption{The evolution of the loss on the training and validation data with respect to the training epoch. The loss of the validation data plateaus out earlier than the training loss, 
but ceases to improve around 70 epochs. It instead becomes worse than the loss of training,
triggering \texttt{callbacks} to stop the training process.
\label{Fig:loss}}
\end{center}
\end{figure}

To prevent overfitting to the training data, we resort to \texttt{callbacks}, which can cease the training process when the loss on the validation data stops improving and save the layer weights with the best performance on the validation set.
Overfitting refers to the phenomenon that a trained network performs very well on the data set that it was trained on but fails to generalize to different data. One of the main reasons is that the network is overtrained and picks up information from noise or statistical fluctuations from the data rather than learning the underlying pattern.
The evolution of the loss functions is shown in Fig.~\ref{Fig:loss}.
The loss of the training data drops significantly until 20 epochs or so and levels off around 40 epochs with a much smaller
decreasing rate.
On the other hand, the validation loss begins with a lower value and flattens at a much earlier time.
After roughly 75 epochs, the curve of validation turns upward and catches up with the training one,
a herald of overfitting, and the training soon is stopped by \texttt{callbacks}. 

Before discussing our results, it is worthwhile to point out that,
as we shall see below, there exists intrinsic noise in the MCMC scan results due to the difficulty in finding the true minimum in a high-dimensional parameter space, \textit{e.g.},
the MCMC chain gets stuck in a local minimum, failing to reach the
global minimum.

\section{Results}
\label{sec:result}

\begin{figure*}[t]
  \centering
\includegraphics[width=0.49\textwidth]{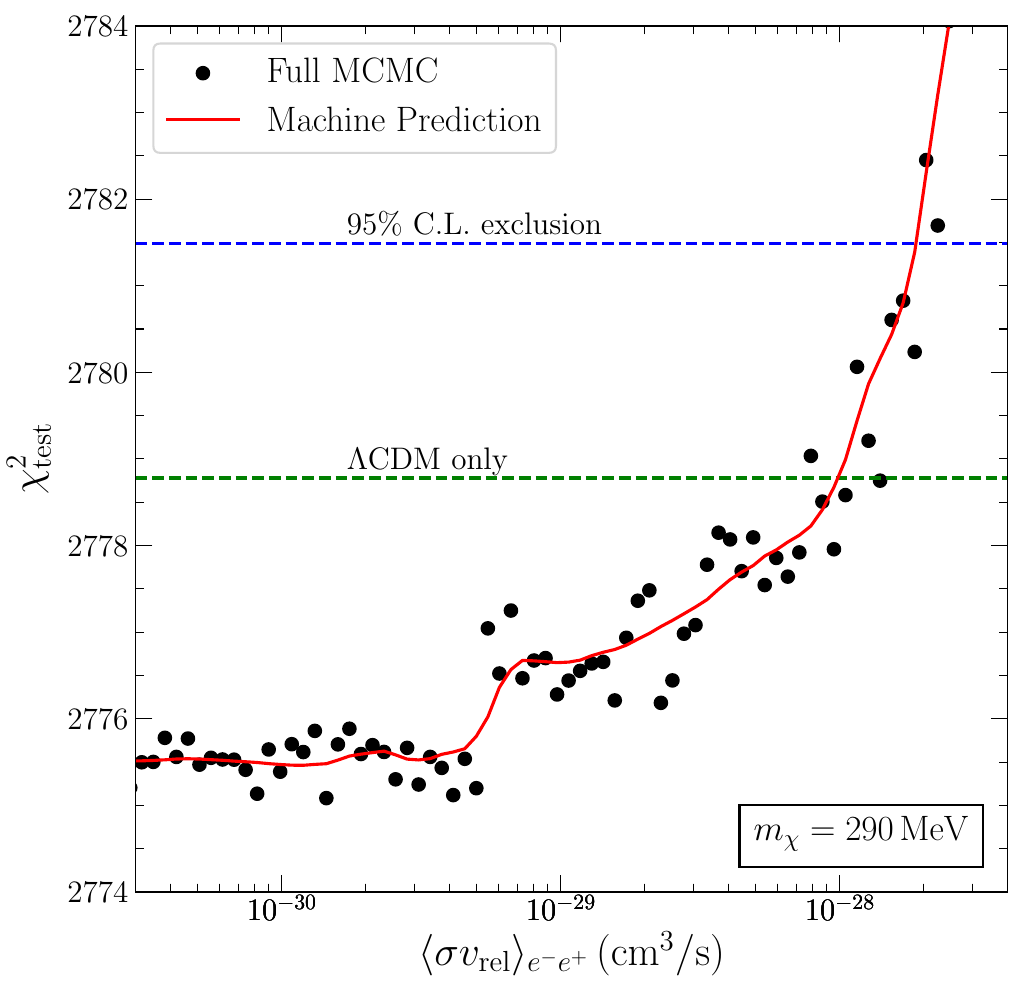}
\includegraphics[width=0.49\textwidth]{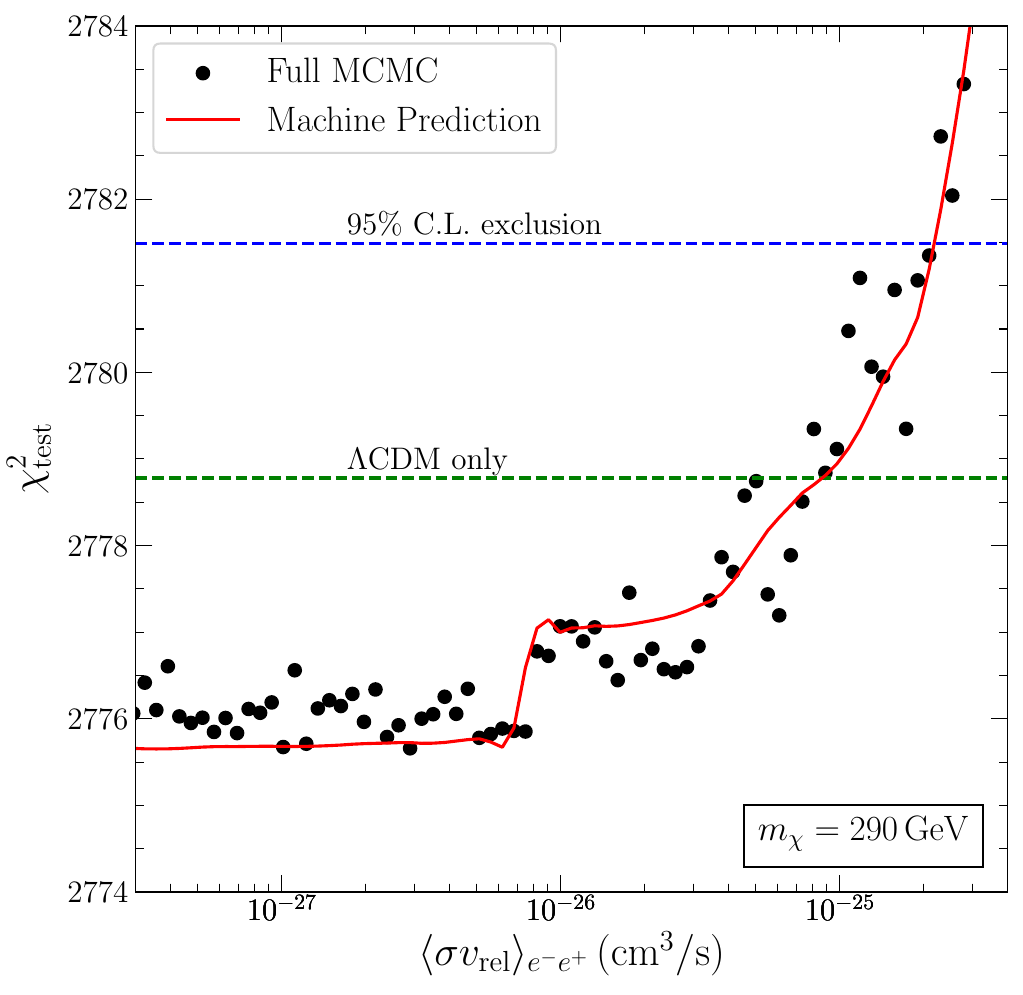}
\caption{The one-dimensional $\chi^2_{\rm test}$ distribution for the $e^- e^+$ channel, from the full MCMC scans (black dot) and the network predictions (red solid curve) for two benchmark values of $m_\chi$, $290$ MeV~(left panel) and $290$ GeV~(right panel). The blue dashed line marks the $95\%$ C.L. exclusion limit on $\sv$ given $m_\chi$. 
}
  \label{Fig:1Dchisq}
\end{figure*}

\begin{figure*}[t]
  \centering
\includegraphics[width=0.49\textwidth]{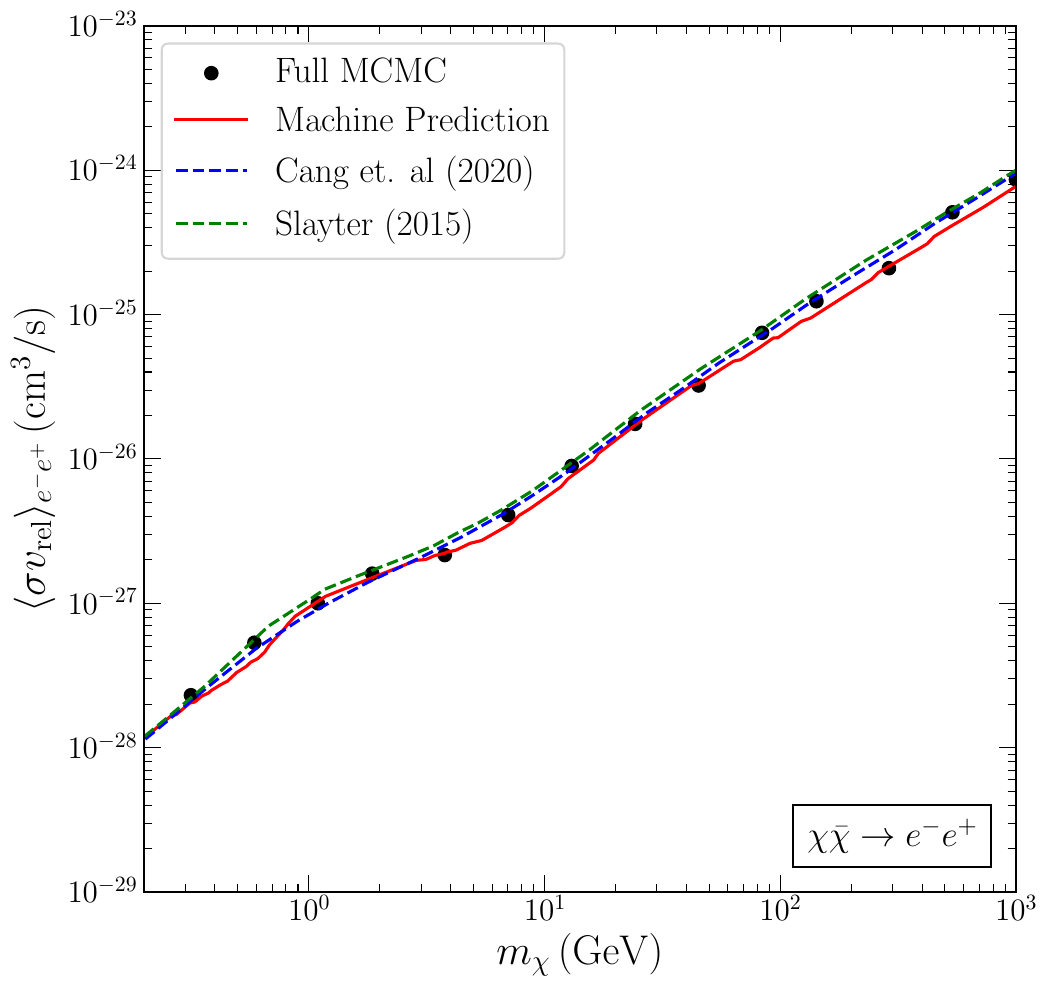}
\includegraphics[width=0.49\textwidth]{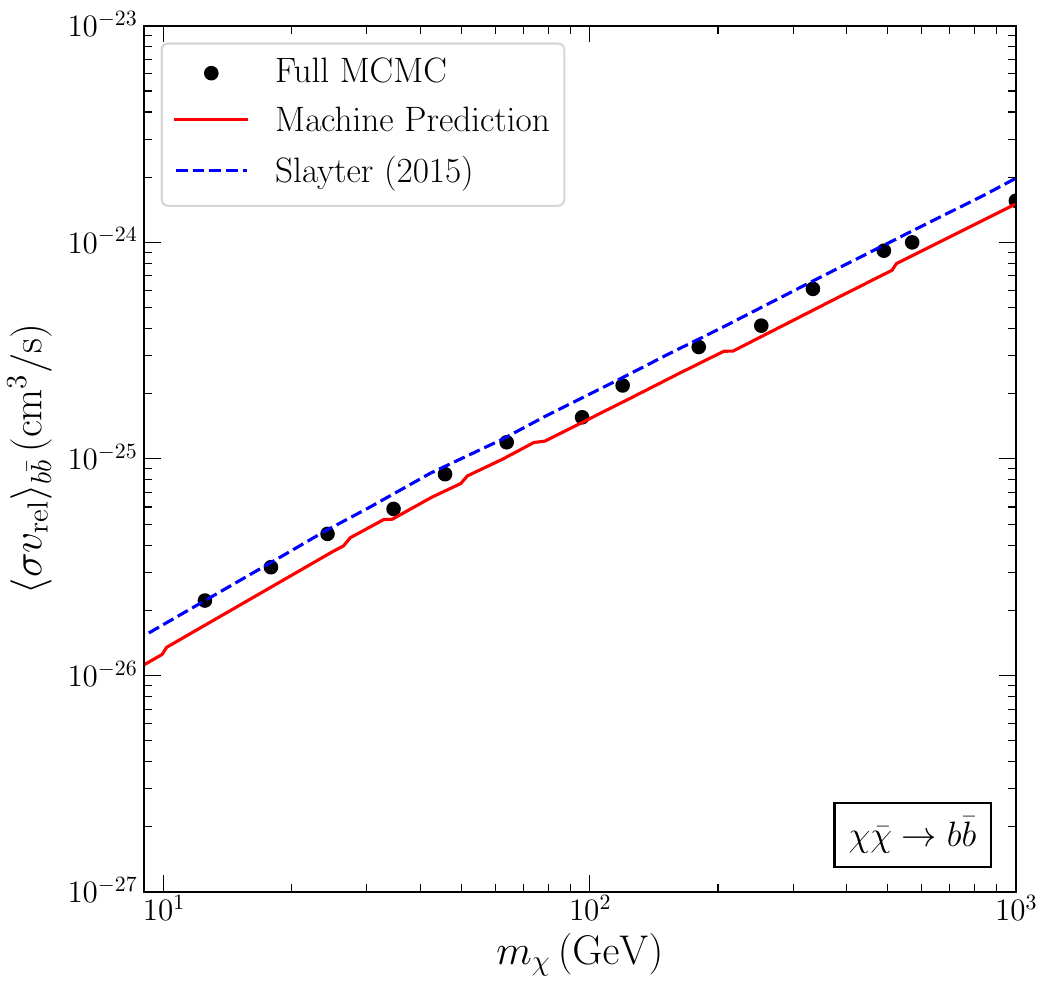}
\caption{Comparison of the $95\%$ C.L. constraints inferred from the full MCMC scans~(black dot) and network predictions~(red solid curve) for $e^- e^+ $~(left panel) and $b\bar{b}$~(right).
We also include some existing bounds from Refs.~\cite{Slatyer:2015jla,Cang:2020exa}. 
One can see that the constraints derived in this work from both the full MCMC scan and the network agree well with the existing bounds.
}
  \label{Fig:fullMC_previous}
\end{figure*}

In this section we present our results and 
demonstrate the predictive power of the network trained on 
the data comprising four annihilation channels: 
$e^- e^+$, $b\bar{b}$, $W^- W^+$ and $\mu^- \mu^+$.
First, we show the one-dimensional $\chi^2_{\rm test}$ distribution for $e^- e^+$ channel with $m_\chi = 290\,{\rm MeV}$ (left panel)  and $m_\chi =290\,{\rm GeV}$ (right panel) in Fig.~\ref{Fig:1Dchisq}. 
The black dot denotes the result of the full MCMC scans~(including both the training and validation datasets), while the red solid line represents the network prediction.
The corresponding $\chi^2_{\rm test}$ for the $\Lambda$CDM-only scenario (green dashed line) and 95$\%$ C.L. exclusion limit (blue dashed line) are also shown for comparison.
It is noticeable that the value of $\chi^2_{\rm test}$ fluctuates among adjacent data points due to the intricacy of minimization in high-dimensional space that leads to  
the innate noise mentioned above.
In this case, by minimizing the loss function, the network manages to find general, overall correlation and trend out of the training data
and naturally yields much smoother prediction curves, as shown in Fig.~\ref{Fig:1Dchisq}.

When the annihilation cross-section is small, the existence of the DM-induced energy injection helps fit better to the CMB observables\footnote{We have also confirmed that $\chi^2_{\rm test}$ will eventually increase and become equal to $\chi^2_{\Lambda\rm{CDM}}$ if one keeps decreasing the cross-section.}. 
On the other hand, DM contributions gradually become important compared to those from the background when $\langle \sigma v_{\rm rel} \rangle$ is large enough, with the increment in $\chi^2_{\rm test}$ proportional to $\langle \sigma v_{\rm rel} \rangle$.
The network successfully captures the overall correlation between $\langle \sigma v_{\rm rel} \rangle$ and $\chi^2_{\rm test}$  with the smoother interpolation, diminishing the noise in the data.
The predicted exclusion limits for both light and heavy DM are fairly close to those of the scans, as can be seen from Fig.~\ref{Fig:1Dchisq}.

In Fig.~\ref{Fig:fullMC_previous}, we show  95$\%$ C.L. constraints on channels of $e^- e^+$ and $b\bar{b}$ derived from the full MCMC scans (black dots) and network predictions (red solid curves).
These channels are selected as they have distinct energy spectra from DM annihilation, while the spectra of $W^- W^+$ and $\mu^- \mu^+$ are quantitatively similar to that of $b \bar{b}$.
Some of the existing bounds in the literature (dashed curves) are also included for comparison: the constraints from Ref.~\cite{Slatyer:2015jla}, that is based on Planck 2015 data, for both $e^- e^+$ and $b \bar{b}$, and the limit derived in Ref.~\cite{Cang:2020exa}, based on Planck 2018 data, for $e^- e^+$.
Note that for the $b \bar{b}$ channel we require $m_\chi \geq 9\,{\rm GeV}$ to avoid the threshold effect for the DM mass around $5\,{\rm GeV}$, when generating $dN_{e,\gamma}^F/dE$.
Three comments are in order as follows. First, the constraints derived from the MCMC scan results agree very well with the existing ones, reinforcing the validity of our implementation of DM-induced contributions into \texttt{CLASS} and \texttt{MontePython}. Second, consistency between the MCMC results and network predictions corroborates again that the trained network successfully learns the underlying patterns from the training data. Therefore, it can be used as an efficient estimator of chi-square values and a satisfactory method of deriving exclusion limits on DM annihilation into SM particles.
Finally, the discrepancy between inferred constraints from Planck 2015 and 2018 data is quite small, and hence the constraints based on different data set are consistent.

\begin{figure*}[t]
  \centering
\includegraphics[width=0.49\textwidth]{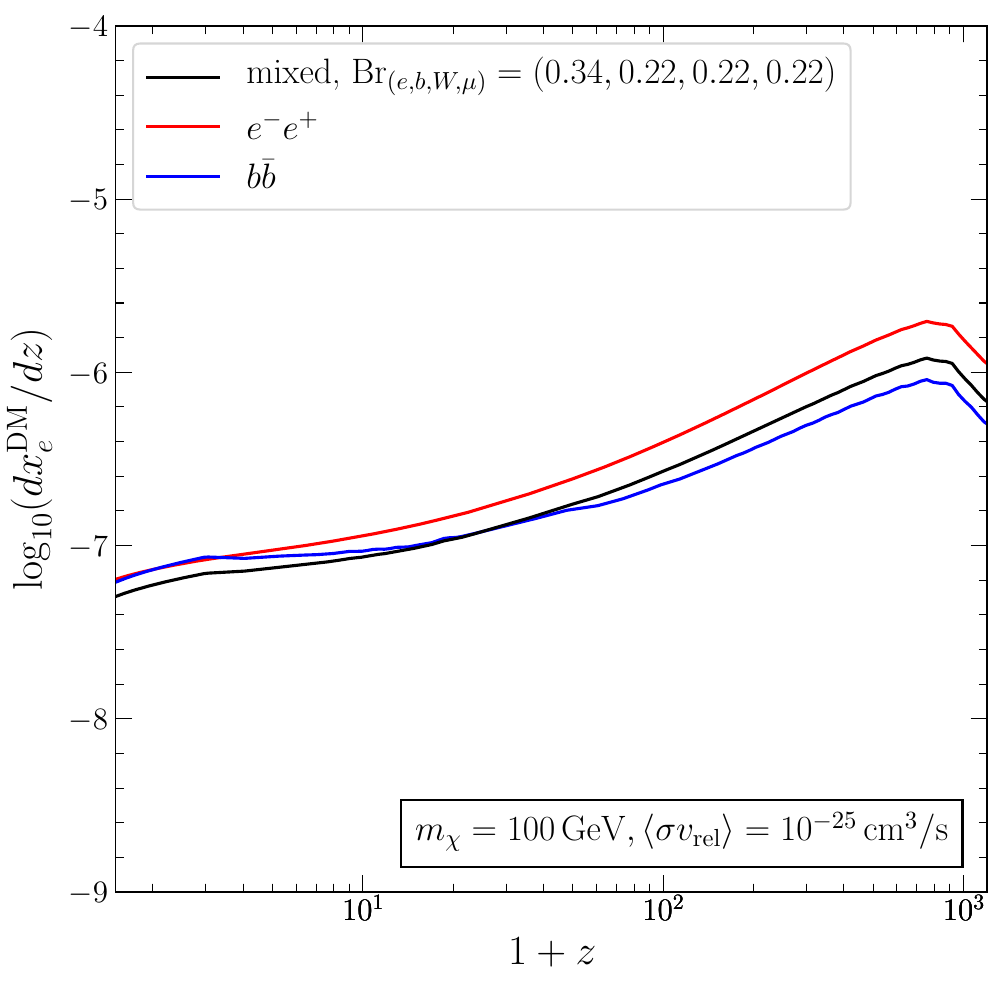}
\includegraphics[width=0.49\textwidth]{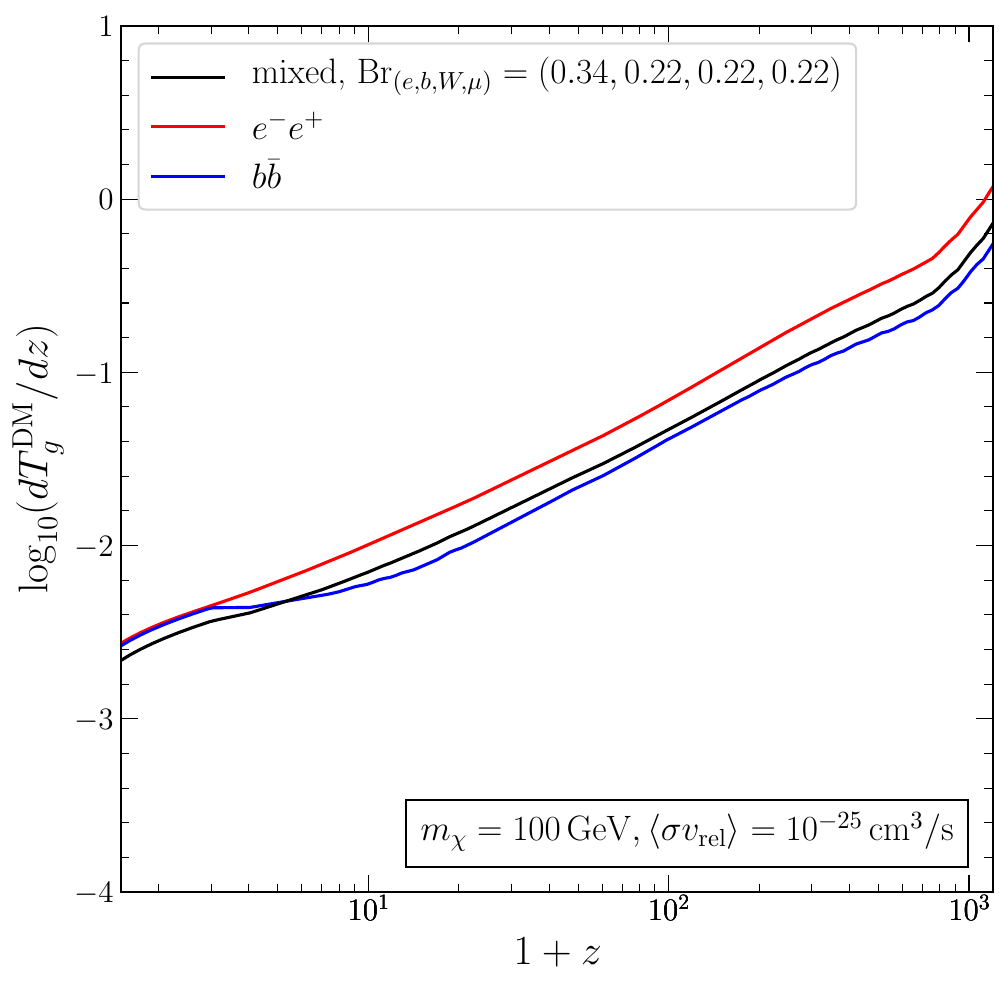}
\caption{Left panel: comparison of the input vector $dx_e^{\rm DM}/dz$ for $e^- e^+$ (red), $b\bar{b}$ (blue), and the mixed channel (black) of branching ratios ${\rm Br}_{(e,b ,W,\mu)} = (0.34, 0.22, 0.22,  0.22)$. We take $m_\chi = 100\,{\rm GeV}$ and $\sv = 10^{-25}\,{\rm cm}^3/{\rm s}$. Right panel: comparison of the input vector $dT_g^{\rm DM}/dz$ with same setting as the left panel.  
}
\label{Fig:inputvec}
\end{figure*}

\begin{figure*}[t]
  \centering
\includegraphics[width=0.55\textwidth]{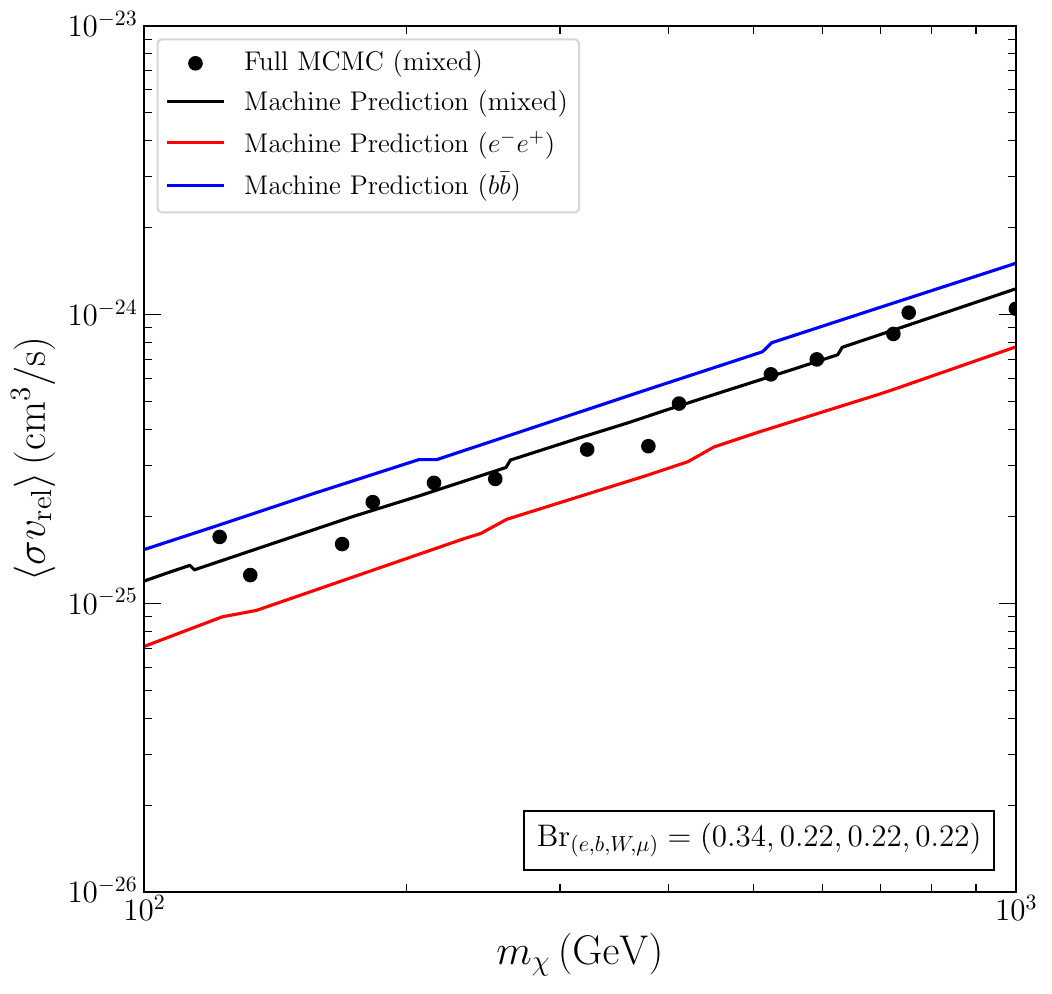}
\caption{Comparison of the $95\%$ C.L. constraints inferred from the full MCMC scans (black dot) and network predictions (black curve) for the mixed channel of  branching ratios ${\rm Br}_{(e,b ,W,\mu)} = (0.34, 0.22, 0.22,  0.22)$.
It demonstrates the predictive power of the network on the unseen data.
}
\label{Fig:mixedchannel}
\end{figure*}

After demonstrating how well the trained network can decipher the data, on part of which it has been trained.
We take one step further to appraise the network performance on unseen data generated from a different mixed channel with the branching ratios of ${\rm Br}_{(e,b ,W,\mu)} = (0.34, 0.22, 0.22,  0.22)$.
The comparisons of the input vectors for neural networks, $dx_e^{\rm DM}/dz$ and $dT_g^{\rm DM}/dz$, of $e^- e^+$ (red), $b \bar{b}$ (blue) and the mixed channel (black) are displayed in Fig.~\ref{Fig:inputvec}.
From Fig.~\ref{Fig:mixedchannel}, it is very intriguing to see that the network which only sees data from four independent channels yields relatively precise, consistent predictions~(black curve) on the newly mixed channel compared to the MCMC scans~(black dots). The 95$\%$ confidence limit for the mixed channel lies between those of the $b \bar{b}$ and $e^- e^+$ channels as expected~(recall the bounds on the $W^- W^+$ and $\mu^- \mu^+$ channels are similar to $b \bar{b}$ channel due to the similar energy spectra); see also Fig.~\ref{Fig:inputvec}.
There also exist wiggles on the curves of prediction that originate from the noise on the training data.
In this case, we have demonstrated that the trained network is capable of producing quantitatively correct exclusion limits on a channel with arbitrary branching ratios. It is plausible that similar performance can also be attained in completely new annihilation channels such as $\tau^-\tau^+$ or $u\bar{u}$, as their energy spectra are not too different from the four channels considered here.
Moreover, to infer the exclusion limit on the mixed channel, we generated 595 data points which take \texttt{MontePython} roughly 10 days when running with 256 CPUs in parallel. On the other hand, it only takes the network less than a minute to predict values of $\delta \chi^2$ for these 595 data points, underscoring  the striking efficiency of the neural network as an estimator of the CMB constraints.

To conclude, the neural network has been proven to be a competent, time-saving method to obtain CMB constraints on DM annihilation. When properly trained with an sufficient amount of data, it delivers consistent results even on unseen data, avoids enormous workload of MCMC scans, and greatly reduces the time needed to attain the results. 
We summarize our results in Table.~\ref{tab:chi_dif}, showing the difference between values of the true and predicted $\delta \chi^2$ on the training and validation data, as well as the relative error of the bound on $\sv$.
\begin{table}
 	\centering
	\begin{tabular}{| P{6cm} | P{3cm}| P{3cm} |}
		\hline
		& $\langle\Delta\rangle$ & $\sigma_{\Delta}$ \\
		\hline
		Training data & -0.0404 &  0.849 \\
		\hline
		Validation data & -0.0540 &  0.850\\
		\hline
		$\sv_{95 \%}$ on four channels & -0.151 &  0.0967\\
		\hline
		$\sv_{95 \%}$ on mixed channel & 0.0162 &  0.124\\
		\hline
	\end{tabular}
	\caption{The summary of the difference between the network predictions~(denoted by $\mathcal{P}$ below) and MCMC~($\mathcal{M}$) data. We quantify the deviation by the mean~(second column) and standard deviation~(third column) of the differences.
	The second and third rows represent the difference on $\delta\chi^2$, \textit{i.e}, $\Delta \equiv 
	(\delta \chi^2)_\mathcal{P} -
	(\delta\chi^2)_\mathcal{M}$ for the training data and validation data.
	The last two rows show the relative error of  the $95 \%$ bound on $\sv$, \textit{i.e.},
	$\Delta \equiv (\sv_\mathcal{P} - \sv_\mathcal{M}) / \sv_\mathcal{P}$. }
	\label{tab:chi_dif}
\end{table}
The discrepancy is partially attributed to the intrinsic fluctuations in the data associated with difficulties of minimization in the high-dimensional space.
Furthermore, degrees of discrepancy are quite similar between the training and validation data, ensuring
the training is not plagued with overfitting.
Overall, we observe good agreement between the network predictions and MCMC results.
The fact that the mean value of the deviation appears
to be smaller for the mixed channel than the independent channels is
attributed to the intrinsic noise in the data and accidental. 
The corresponding standard deviations, however, are quite similar that implies the network generalizes well to the unseen mixed channel.

\section{Conclusions}
\label{sec:conclusion}

In this work, we propose a novel way---a CNN estimator---to infer CMB bounds on
DM-induced energy injection.
We closely conform to the philosophy of model-independence promoted in existing works to bypass or considerably reduce  time-consuming MCMC scans over a large hyperspace space of cosmological parameters and degrees of freedom in DM models.
More importantly, we take full advantage of neural networks' power in discovering underlying patterns of input and output parameters to provide a fast and efficient method of inferring the CMB limits for annihilating DM models with \textit{several} final states and \textit{arbitrary} branching ratios.  

To realize the goal, we select the DM-induced changes of the ionization fraction and gas temperature
as the variables to quantify DM effects on the CMB.
As confirmed by the simplified residual likelihood map, the impacts can be genuinely reflected in terms of the chosen quantities. Next, we use various combinations of the DM mass
and annihilation cross-section from four independent channels, $e^- e^+$, $b\bar{b}$, $\mu^- \mu^+$ and $W^- W^+$, and compute corresponding CMB power spectra which are affected by DM energy injection.
The maximum likelihood for each set of the mass and cross-section can be attained via profiling out the cosmological and nuisance parameters. The dataset containing input parameters~($dx_e^{\rm DM}/dz$ and $dT_g^{\rm DM}/dz$) and the target variable~(likelihood) are split into the training data and validation
data and fed into a CNN for training.

The decent performance on both the training and validation data has been achieved -- the difference in chi-square between the full MCMC scans and network predictions is less than one unit on average.
It takes only a matter of \textit{seconds}, instead of days, for the network to make prediction.
More strikingly, the equally good performance also manifests itself at a new mixed channel where DM annihilates simultaneously into the four final states. In other words, the CNN estimator is proven to be a powerful and efficient tool for inferring CMB bounds on DM models.

This work is a valuable stepping-stone to a fully developed CNN estimator that can constrain \textit{in a matter of seconds} energy injection from any types of DM models or exotic scenarios beyond the standard cosmology based on the CMB measurements.
It can also predict other important cosmological parameters at the same time when including these parameters into the training data.

\acknowledgments

We thank Masahiro Kawasaki, Kazunori Nakayama, and Toyokazu Sekiguchi for providing us the tables and Chien Lin for early collaboration. We are very grateful to Florian Niedermann for extremely helpful discussions on \texttt{CLASS} and \texttt{MontePython}. 
W.-C.~Huang is supported by the Independent Research Fund Denmark, grant number DFF 6108-00623.
J.-L.~Kuo is supported by the Austrian Science Fund
FWF under the Doctoral Program W1252-N27 Particles
and Interactions.
Y.-L.~S.~Tsai was funded by the Ministry of Science and Technology Taiwan 
under Grant No. 109-2112-M-007-022-MY3.
The authors would like to acknowledge that this work was performed using the \href{https://escience.sdu.dk/index.php/ucloud/}{UCloud} computing and storage resources, managed and supported by eScience center at SDU.

 
\appendix

\section{Decay}
\label{Sec:decay}

For DM decay, the energy injection rate is proportional to $n_\chi / \tau_{\rm DM}$, where $\tau_{\rm DM}$ is the lifetime of DM.
The contribution of DM decay to ionization and heating is thus expressed as
\begin{align}
-\left[{\frac{dx_e^{\rm DM}}{dz}}\right]_{\rm decay} &= \sum_{F} {\rm Br}_F\int_z \dfrac{dz'}{H(z')(1+z')} \dfrac{n_\chi(z')}{2 n_{\rm H}(z')\tau_{\rm DM}} \dfrac{m_\chi}{E_{\rm RY}} \dfrac{d\chi^F_i(m_\chi,z,z')}{dz}\,, \nonumber \\
-\left[ \dfrac{dT_g^{\rm DM}}{dz}\right]_{\rm decay} &= \sum_{F} {\rm Br}_F\int_z \dfrac{dz'}{H(z')(1+z')}\dfrac{n_\chi(z')}{3n_{\rm H}(z')\tau_{\rm DM}} m_\chi \dfrac{d\chi^F_h(m_\chi,z,z')}{dz}\,.
\label{eq:dxdz_decay}
\end{align}

\section{$p$-wave annihilation}
\label{Sec:pwave}

The $p$-wave annihilation cross-section $\sv$ is proportional to $v_{\rm rel}^2$, which can be written as
\begin{equation}
\sv = \dfrac{\langle v_{\chi}^2 \rangle}{v_{0}^2} \svnow,
\end{equation}
with DM velocity at any time $v_{\rm \chi}$, a reference velocity $v_0$ and a reference cross-section $\sv_0$ for $p$-wave annihilation when $\langle v_{\chi}^2 \rangle = v_0^2$. 
We follow the convention~\cite{Diamanti:2013bia} and choose $v_0 = 100\,{\rm km/s}$, which is roughly the velocity dispersion of DM in halos at $z=0$, for easy comparison with the result of indirect detection.
After DM kinetic decouples and becomes non-relativistic, its temperature $T_\chi$ redshifts as $(1+z)^2$, thus we can deduce that
\begin{equation}
\label{Eq:vandz}
    \dfrac{\langle v_{\chi}^2 \rangle}{v_{0}^2} = \left(\dfrac{1+z}{1+z_{\rm ref}} \right)^2\,,
\end{equation}
where $z_{\rm ref}$ is the redshift that the root-mean-square velocity $v_{\rm rms} \equiv \sqrt{\langle v_{\chi}^2\rangle} = \sqrt{3 T_\chi / m_\chi }$ of DM equals to $v_0$.
Note that we have assumed DM is an ideal gas and adopted the energy equipartition.
It is useful to express $z_{\rm ref}$ as a function of the DM kinetic decoupling temperature $T_{\rm kd}$.
At DM kinetic decoupling ($T_\chi = T_{\rm kd}$), Eq.~\eqref{Eq:vandz} could be transformed into
\begin{equation}
\label{Eq:zref1}
    1+z_{\rm ref} = v_0 (1+z_{\rm kd}) \sqrt{\dfrac{m_\chi}{3T_{\rm kd}}}\,,
\end{equation}
where $z_{\rm kd}$ is the redshift of DM kinetic decoupling. 
Furthermore, $z_{\rm kd}$ can be written in terms of current CMB temperature $T_{\gamma,0} = 2.35 \times 10^{-10}\,{\rm MeV}$ as 
\begin{equation}
\label{Eq:zkd}
1 + z_{\rm kd} = \dfrac{T_{\rm kd}}{T_{\gamma,0}} \simeq 4.26\times 10^9 \left(\dfrac{T_{\rm kd}}{{\rm MeV}} \right)\,.
\end{equation}
Putting together Eq.~\eqref{Eq:zref1} and Eq.~\eqref{Eq:zkd}, we obtain 
\begin{equation}
     1+z_{\rm ref} = 2.59 \times 10^7 \left(\dfrac{T_{\rm kd}}{{\rm MeV}} \right)^{1/2} \left(\dfrac{m_\chi}{{\rm GeV}} \right)^{1/2}\,.
\end{equation}

Finally, the thermal-averaged $p$-wave annihilation cross-section can be expressed as
\begin{equation}
    \sv = \left( \dfrac{1+z}{1+z_{\rm ref}} \right)^2 \svnow\,.
\end{equation}
Similar to $s$-wave annihilation, we can write the DM contribution to ionization and heating as 
\begin{align}
\label{Eq:energy_injection_p-wave}
-\left[\dfrac{dx_e}{dz}\right]_{p\text{-}{\rm  wave}} &= \sum_{F}  {\rm Br}_F
              \int_z \dfrac{dz'(1+z') }{H(z') (1+z_{\rm ref})^2} 
       \dfrac{n^2_\chi(z') \svnow}{2n_{\rm H} (z')} 
       \dfrac{m_\chi}{E_{\rm RY}} \dfrac{d\chi^F_i(m_\chi ,z,z')}{dz}\,, \nonumber \\
-\left[\dfrac{dT_b}{dz} \right]_{p\text{-}{\rm wave}} &= \sum_{F} {\rm Br}_F
       \int_z \dfrac{dz'(1+z') }{H(z')(1+z_{\rm ref})^2} 
       \dfrac{n^2_\chi(z') \svnow}{3n_{\rm H}(z')}m_\chi \dfrac{d\chi^F_h(m_\chi,z,z')}{dz}\,.
\end{align}

\section{Breit-Wigner enhancement}
\label{Sec:resonance}

In the Breit-Wigner enhancement scenario, DM annihilates to SM particles via a narrow resonance.
Following Ref.~\cite{Ibe:2008ye}, we consider a scalar resonance $\phi$ and 
the general DM annihilation cross-section via the resonance ($\chi \bar{\chi}\to \phi \to f \bar{f}$) can be written as
\begin{equation}
\label{Eq:xsec_general}
    \sigma= \dfrac{16\pi}{E_{\rm cm}^2 \bar{\beta}_i \beta_i} \dfrac{m_\phi^2 \Gamma_\phi^2}{(E_{\rm cm}^2 - m_\phi^2)^2 + m_\phi^2 \Gamma_\phi^2} \times B_i B_f\,,
\end{equation}
where $E_{\rm cm}$ is the center-of-mass energy, $m_\phi$ and $\Gamma_\phi$ are the mass and decay rate of the resonance $\phi$. 
The initial state and final state space phase factors are $\Bar{\beta_i} = \sqrt{1-4\mchi^2/m_\phi^2}$ evaluated at the resonance 
and ${\beta_i} = \sqrt{1-4\mchi^2/E^2_{\rm cm}}$ at $E_{\rm cm}$ of the collision.
The branching ratio of $\phi$ decaying to $\chi \bar{\chi}$ and $f\bar{f}$ are 
$B_\chi$ and $B_f$, respectively. 

Considering DM are non-relativistic at the recombination epoch and later times, we are allowed to adopt the Maxwell-Boltzmann velocity distribution for DM and use the Gaussian average to compute the  thermal-averaged annihilation cross-section, which reads
\begin{equation}
\label{Eq:Gaussian_integral}
     \sv  = \dfrac{1}{(2\pi v_{\rm rms}^2 /3)^3 } \int d \vec{v}_\chi \int d \vec{v}_{\bar{\chi}} \, e^{-3(v_\chi^2 + v_{\bar{\chi}}^2)/(2 v_{\rm rms}^2)} \times \sigma v_{\rm rel}\,,
\end{equation}
where $\vec{v}_{\chi, \bar{\chi}}$ are the velocities of initial states and $|\vec{v}_{\chi, \bar{\chi}}|= v_{\chi, \bar{\chi}}$.
In the non-relativistic limit ($v_\chi, v_{\bar{\chi}} \ll 1$ and $E_{\rm cm}^2 \sim 4 m_\chi^2$), the center-of-mass energy can be expanded as $E_{\rm cm}^2 = 4m_\chi^2 + m_\chi^2 v_{\rm rel}^2$, and the relative velocity can be approximated as $v_{\rm rel } \simeq 2 \beta_i$.

The condition of annihilation near a narrow resonance is fulfilled when 
\begin{equation}
\label{Eq:narrow_condition}
    m_\phi^2 = 4 m_\chi^2 (1- \delta), |\delta| \ll 1\,,
\end{equation}
Note that positive~(negative) $\delta$ will imply DM annihilates below~(above) the pole in Eq.~\eqref{Eq:xsec_general} when $\bar{\beta}_i = 0$  (when $m_\phi = 2m_\chi$).
The pole becomes unphysical when $\delta > 0$ ($m_\phi < 2m_\chi$), but it can be regarded as analytic continuations of those quantities from the physical region as $\delta < 0$. 
With Eq.~\eqref{Eq:narrow_condition} and defining $\gamma \equiv \Gamma_\phi / m_\phi$, we can write Eq.~\eqref{Eq:xsec_general} as 
\begin{equation}
    \sigma = \dfrac{16\pi}{m_\phi^2 \bar{\beta}_i \beta_i} \dfrac{\gamma^2}{(\delta+ v_{\rm rel}^2 /4)^2 +\gamma^2} \times B_i B_f\,.
\end{equation}
Ref.~\cite{Ibe:2008ye} provides us a good approximation for the Gaussian average, 
\begin{equation}
    \sv \simeq \dfrac{32\pi}{m_\phi^2 \bar{\beta}_i} \dfrac{\gamma^2}{[\delta + v_{\rm rms}^2 / (3\sqrt{2})]^2 + \gamma^2} \times B_i B_f\,,
\end{equation}
applicable for the parameter region $v_{\rm rms} \ll 1$ and $\delta < 0$ that we will consider. 
%
For simplicity, we reparameterize $\sv$ as 
\begin{equation}
    \sv = \dfrac{\delta^2 + \gamma^2}{[\delta + v_{\rm rms}^2 / (3\sqrt{2})]^2 + \gamma^2} \sv_0\,,
\end{equation}
where $\sv_0$ for the Breit-Wigner enhancement is defined as the thermal-averaged annihilation cross-section at $T_\chi = 0$ ($v_{\rm rms } = 0 $), written as 
\begin{equation}
\sv_0 = \dfrac{32\pi B_i B_f}{m_\phi^2 \bar{\beta}_i} \dfrac{\gamma^2}{\delta^2 + \gamma^2}\,.
\end{equation}

Therefore, for the Breit-Wigner enhancement, the DM contribution to ionization and heating is formulated as
\begin{align}
-\left[{\frac{dx_e}{dz}}\right]_{\rm BW} &= \sum_{F} {\rm Br}_F\int_z \dfrac{dz'}{H(z')(1+z')} 
\dfrac{n^2_\chi(z'){\sv}_0}{2n_{\rm H}(z')} 
\frac{\gamma^2+\delta^2}{[\delta+ v_{\rm rms}^2/(3\sqrt{2})]^2+\gamma^2}\dfrac{m_\chi}{E_{\rm RY}} 
\dfrac{d\chi^F_i(m_\chi,z,z')}{dz}\,, \nonumber \\
-\left[ \dfrac{dT_g}{dz}\right]_{\rm BW} &= 
\sum_{F} {\rm Br}_F\int_z \dfrac{dz'}{H(z')(1+z')}\dfrac{n^2_\chi(z'){\sv}_0}{3n_{\rm H}(z')}
\frac{\delta^2+\gamma^2}{[\delta + v_{\rm rms}^2/(3\sqrt{2})]^2+\gamma^2} m_\chi \dfrac{d\chi^F_h(m_\chi,z,z')}{dz}\,.
\label{Eq:energy_injection_BWE}
\end{align}
The redshift dependence of $v_{\rm rms}$ can be derived as 
\begin{equation}
    v_{\rm rms} (z) = \sqrt{\dfrac{T_\chi (z)}{T_{\rm kd}}} = \dfrac{1+z}{1+z_{\rm kd}} \simeq 2.3 \times 10^{-10} (1+z) \left(\dfrac{T_{\rm kd}}{{\rm MeV}} \right)^{-1}\,,
\end{equation}
where Eq.~\eqref{Eq:zkd} is used.

\bibliography{ref}

\end{document}